\documentclass[10pt, conference, letterpaper]{IEEEtran}

\usepackage{cite}
\usepackage[pdftex]{graphicx}
\usepackage{amsmath}
\usepackage{amsfonts}
\usepackage{amssymb}
\usepackage{subcaption}
\usepackage{caption}
\usepackage{bm}
\usepackage{comment}
\usepackage{xcolor}
\usepackage{algorithm} 
\usepackage{algpseudocode}
\usepackage{multirow}
\usepackage{booktabs}
\usepackage[printonlyused]{acronym}
\newcommand{\bigO}{\mathcal{O}}

\newcommand{\probP}{\text{I\kern-0.15em P}}

\newcommand{\blue}{\textcolor{black}}

\begin{document}


\title{A Flexible Multi-Agent Deep Reinforcement Learning Framework for Dynamic Routing and Scheduling of Latency-Critical Services}

\author{\IEEEauthorblockN{Vincenzo Norman Vitale}
\IEEEauthorblockA{
\textit{\small University of Naples Federico II}\\
\small Naples, Italy\\
vincenzonorman.vitale@unina.it}
\and
\IEEEauthorblockN{Antonia Maria Tulino}
\IEEEauthorblockA{\textit{\small University of Naples Federico II}\\
\small Naples, Italy\\
antoniamaria.tulino@unina.it}
\and
\IEEEauthorblockN{Andreas F. Molisch}
\IEEEauthorblockA{
\textit{\small University of Southern California}\\
\small Los Angeles, USA \\
molisch@usc.edu}
\and
\IEEEauthorblockN{Jaime Llorca}
\IEEEauthorblockA{
\textit{\small University of Trento}\\
\small Trento, Italy\\
jaime.llorca@unitn.it}
}


%

\IEEEoverridecommandlockouts
\IEEEpubid{
	\begin{minipage}{3\columnwidth}
		\centering
		{\footnotesize
			\vspace{50pt}
			This work has been submitted to the IEEE for possible publication. Copyright may be transferred without notice, after which this version may no longer be accessible.
		}
	\end{minipage}
}

\maketitle

\begin{abstract}
Timely delivery of delay-sensitive information over dynamic, heterogeneous networks is increasingly essential for a range of interactive applications, such as industrial automation, self-driving vehicles, and augmented reality. However, most existing network control solutions target only average delay performance, falling short of providing strict End-to-End (E2E) peak latency guarantees. This paper addresses the challenge of reliably delivering packets within application-imposed deadlines by leveraging recent advancements in Multi-Agent Deep Reinforcement Learning (MA-DRL). After introducing the Delay-Constrained Maximum-Throughput (DCMT) dynamic network control problem, and highlighting the limitations of current solutions, we present a novel MA-DRL network control framework that leverages a centralized routing and distributed scheduling architecture. The proposed framework leverages critical networking domain knowledge for the design of effective MA-DRL strategies based on the Multi-Agent Deep Deterministic Policy Gradient (MADDPG) technique, where centralized routing and distributed scheduling agents dynamically assign paths and schedule packet transmissions according to packet lifetimes, thereby maximizing on-time packet delivery. The generality of the proposed framework allows integrating both data-driven \blue{Deep Reinforcement Learning (DRL)} agents and traditional rule-based policies in order to strike the right balance between performance and learning complexity. Our results confirm the superiority of the proposed framework with respect to traditional stochastic optimization-based approaches and provide key insights into the role and interplay between data-driven DRL agents and new rule-based policies for both efficient and high-performance control of latency-critical services.

\end{abstract}

\IEEEpeerreviewmaketitle

\section{Introduction}
The traditional divide between computation-centric systems and communication-centric networks is rapidly fading with the advent of NextG networks. Driven by advancements in softwarization, programmability, and edge-to-cloud integration, these networks are transforming into interconnected, compute-integrated ecosystems. They are designed to support a new generation of Real-Time Interactive (RTI) services, which encompass resource-intensive, time-sensitive applications across a range of fields. From cyber-physical systems like industrial automation and smart city management to human-centered applications such as telemedicine and extended reality, these services impose unique and stringent latency requirements on network control~\cite{cai2022compute}.

Meeting these rigorous demands has made E2E service orchestration a central focus in NextG networks. Such an orchestration problem addresses three interrelated challenges: service placement, flow routing, and resource allocation, all bound by the Quality of Service (QoS) requirements specific to each application and additional context-specific constraints.  The Cloud Network Flow (CNF) framework provides a unified approach for tackling these interconnected requirements, jointly formulating placement, routing, and resource allocation problems as single flow problem on a properly augmented graph~\cite{barcelo2016iot}. A number of CNF-based algorithms have been developed in order to provide efficient polynomial-time solutions to the long-term centralized optimization version of the service orchestration problem (e.g., \cite{Poularakis2020Mobihoc,mauro2024end}). 
To enable comprehensive E2E service orchestration, such centralized long-term optimization policies are complemented with distributed dynamic control policies that adjust routing, scheduling, and resource allocation decisions in response to dynamic changes in network conditions and service demands~\cite{pagliuca2024dual}. 

In the context of dynamic network control, the dynamic packet routing and scheduling problem under bounded average delays is a well-studied problem. 
Algorithms such as the Backpressure (BP) routing algorithm~\cite{tassiulas1990stability} leverage Lyapunov-based stochastic optimization methods to provide fully distributed routing and scheduling solutions with throughput-optimality guarantees.
The Lyapunov drift-plus-penalty (LDP) algorithm, an extension of BP, further supports cost minimization while maintaining throughput optimality~\cite{neely2022stochastic}.

Despite achieving throughput optimality and ensuring bounded average delay, BP and LDP algorithms frequently result in high average delays caused by excessively long and even cyclic routes. To address this shortcoming, algorithms that combine centralized routing with distributed scheduling—such as Universal Max-Weight (UMW) or Universal Cloud Network Control (UCNC) \cite{sinha2017optimal, zhang2021optimal}—can improve average delay performance by selecting acyclic routes while maintaining throughput optimality. Yet, these approaches only address average delay performance and lack guarantees on the exact time a packet will remain in the network before delivery.

In many critical applications, packets that fail to reach their destination within a set deadline—known as their lifetime—are considered outdated and irrelevant \cite{singh2018throughput}. 
Dynamic network control problems with strict packet lifetime constraints hence introduce new challenges that break the suitability of stochastic optimization based approaches that leverage queue stability to guarantee bounded average delays~\cite{cai2022ultra}.
To this end, in this paper, we formulate the Delay-Constrained Maximum-Throughput (DCMT) dynamic network control problem as a Markov Decision Process (MDP), where routing and scheduling actions are taken based on lifetime-based queuing system observations, and provide an efficient Reinforcement Learning (RL) based centralized routing and distributed scheduling solution.

RL-based approaches have shown promise for packet routing and scheduling in networks modeled as MDPs~\cite{boyan1993packet, mammeri2019reinforcement,littman1994markov,luong2019applications}. These methods typically focus on maintaining queue stability \cite{liu2022rl}, reducing average delivery times \cite{you2020toward}, or managing network resources \cite{liu2021constraint}. However, RL-based solutions for packet routing under strict deadline constraints remain an open research problem.
In addition, using RL in heterogeneous communication networks brings crucial challenges related to the dimensionality of state and action representation spaces 
\cite{luong2019applications}.

Our main contributions in this area are:

\begin{itemize}
    \item We model the network control (routing and scheduling) problem with packet lifetime constraints, referred to as the DCMT network control problem, as an MDP, where an optimal policy for maximizing timely throughput can be derived using Deep Reinforcement Learning (DRL).
    \item We propose a flexible multi-agent DRL (MADRL) based network control framework that allows a centralized routing agent and multiple distributed scheduling agents to collaboratively learn routing and scheduling policies that maximize timely throughput.
    \item The proposed strategies progressively leverage multiple levels of networking domain knowledge to relieve RL agents from decisions that can be effectively handled by new rule-based control policies.
    \item Through incremental improvements in strategy design, we provide key insights and sensitivity analysis that allow identifying the parts of the agents' state and action spaces most relevant to improving performance.
\end{itemize}

The remainder of this work is organized as follows. In Section \ref{sec:SOTA}, we provide an overview of Multi-Agent Deep Reinforcement Learning (MADRL) as well as related work using MARL in communication networks. Then, Section \ref{sec:system_model} reports the formal system model, used in Section \ref{sec:DCMT_Problem_Formulation} to provide the DMCT network control problem formulation. Section \ref{sec:framework} introduces the proposed MADRL framework, and Section \ref{sec:network_control_strategies} the associated network control strategies. Section \ref{sec:expermiental_setting_and_results} closes with some considerations regarding the design choices, computational complexity, and possible alternative developments. Section \ref{sec:numerical_results} reports the results of our numerical experiments on the proposed strategies compared to the state-of-art. This is followed by Section \ref{sec:discussion_and_future_dev}, which discusses improvements, open problems, and future developments. Concluding remarks are provided in Section \ref{sec:conclusions}.

\section{Background and Related Work}
\label{sec:SOTA}


\blue{This section provides an overview of the essential background and related literature. In Section \ref{sec:sota_marl}, we present a theoretical overview of Multi-Agent Reinforcement Learning (MARL), covering its fundamental concepts and principles. In Section \ref{sec:sota_MARL_COMM}, we review studies that have applied MARL to the networking domain, thereby establishing the context for our contribution.}
\subsection{Multi-Agent Reinforcement Learning (MARL)}
\label{sec:sota_marl}

Reinforcement learning models the interaction between a learning agent and its environment using the framework of a Markov Decision Process (MDP) \cite{bellman1957markovian}. In this framework, interactions are defined by states, actions, and rewards. Specifically, an agent operates in an environment where, at each time instant \( t \), the state \( s(t) \in \mathcal{S} \) is used to select an action \( a(t) \in \mathcal{A} \) according to its stochastic policy \( \pi(a(t) \mid s(t)) \). Executing this action generates a scalar reward \( r(t) = R(s(t), a(t)) \) and causes the environment to transition to a new state \( s(t+1) \), determined by the probability distribution \( P(s(t+1) \mid s(t), a(t)) \). At \( t = 0 \), the initial state \( \blue{s(0)} \) is sampled from an initial distribution \( \sigma \).

For any given  time interval $[0,T]$,  this process produces a {\em trajectory} (i.e., a sequence) of states, actions, and rewards:
\[
\tau = \left \{s(t), a(t), r(t) \right\}_{t=0}^{T} 
\]
where \(T\) is the trajectory length. The agent's objective is to find an optimal policy \(\pi^*\) that maximizes the total expected return:
\[
R(\tau) = \sum_{t=0}^{T} \gamma^t r_t,
\]
where \(\gamma\) is a discount factor that reduces the value of future rewards.

\blue{The goal of reinforcement learning is to iteratively improve the policy \(\pi\) through interactions with the environment, so that it maximizes the expected cumulative reward over time. This process enables the agent to learn an optimal behavior or strategy for decision-making in complex and uncertain environments.}

The literature also analyzes settings where the agent does not observe the full environment state but instead receives a partial observation \(o = \omega(s)\). This scenario is referred to as a Partially Observed MDP (POMDP) \cite{kaelbling1998planning}. In this case, actions are selected based on the partial observation, following a policy \(\pi(a | o)\).

\subsubsection*{Value Functions}
Many algorithms for reinforcement learning make use of value functions that represent the "goodness" of a state or an action in terms of potential future rewards \cite{sutton2018reinforcement}. A state-value function represents the total expected return from a state \(s\) using policy \(\pi\):

\begin{equation}
    V^{\pi}(s)=\underset{\tau \sim \pi}{\mathbb{E}}\left[R(\tau) \mid s_{0}=s\right],
\end{equation}
where the expectation is over the trajectories induced by policy \(\pi\). Similarly, an action-value function or \(Q\)-function represents the total expected return when choosing action \(a\) in state \(s\):

\[
Q^{\pi}(a,s)=\underset{\tau \sim \pi}{\mathbb{E}}\left[R(\tau) \mid a_{0}=a, s_{0}=s \right].
\]

The optimal action-value function, \(Q^{*}(a,s)\), represents the maximum expected return that can be achieved from state \(s\) by taking action \(a\) and subsequently following the best possible policy:
\[
Q^{*}(a,s) = \max_{\pi} Q^{\pi}(a,s).
\]

This function can be computed using temporal-difference learning \cite{sutton2018reinforcement,watkins1992daya}, an iterative method that updates the \(Q\)-function based on the Bellman optimality equation. The Bellman equation expresses the recursive relationship between the optimal \(Q\)-values:
\[
Q^{*}(a,s) = \mathbb{E}_{s^{\prime} \sim P} \left[ R(a,s) + \gamma \max_{a^{\prime}} Q^{*}(a^{\prime},s^{\prime}) \right],
\]
where \(R(s, a)\) is the immediate reward for taking action \(a\) in state \(s\), \(\gamma\) is the discount factor, and the expectation accounts for the stochasticity of state transitions governed by \(P(s^{\prime} \mid s, a)\).

Once \(Q^{*}(a,s)\) is determined, the optimal policy \(\pi^{*}\) can be directly derived by choosing the action that maximizes \(Q^{*}\) for each state \(s\):
\[
\pi^{*}(s) = \underset{a}{\operatorname{argmax}} \, Q^{*}(a,s).
\]

This means the optimal policy is simply the action-selection rule that always picks the action with the highest expected return according to the optimal \(Q\)-function.

\subsubsection*{\blue{Deep Deterministic Policy Gradient (DDPG)}}
\blue{In policy gradient methods for reinforcement learning~\cite{sutton1999policy}, 
policies are parameterized by a vector \(\theta \in \mathbb{R}^\kappa\), 
where \(\kappa\) denotes the dimension of \(\theta\).} 
Policy gradient methods adjust the parameters \(\theta\) of a policy \(\pi_{\theta}\) directly to maximize the objective
\[
J(\theta) = \underset{\tau \sim \pi_{\theta}}{\mathbb{E}}[R(\tau)],
\]
\blue{where \(R(\tau)\) denotes the total return of a trajectory \(\tau\), 
\(\tau \sim \pi_{\theta}\) indicates that trajectories are sampled according to the policy, 
and \(J(\theta)\) is the expected return over all such trajectories.}

The Policy Gradient Theorem, introduced by Sutton et al.~\cite{sutton1999policy}, expresses the gradient of the objective function \( J(\theta) \) as:
\[
\nabla_{\theta} J(\theta) = \mathbb{E}_{\tau \sim \pi_{\theta}} \left[ \sum_{t=0}^{T} \nabla_{\theta} \log \pi_{\theta}(a_t \mid s_t) \, Q^{\pi}(s_t, a_t) \right].
\]

The term \( \nabla_{\theta} \log \pi_{\theta}(a_t | s_t) \) is known as the \textit{score function} or the gradient of the log-likelihood. It indicates how the policy's parameters should be adjusted to increase the probability of selecting action \( a_t \) in state \( s_t \). By employing the Policy Gradient Theorem, one can perform gradient ascent on the expected return \( J(\theta) \) to optimize the policy parameters \( \theta \).

Building upon the Policy Gradient Theorem, the Deterministic Policy Gradient (DPG) algorithm, introduced by Silver et al.~(2014), extends policy gradient methods to deterministic policies in continuous action spaces.

In DPG, the policy is defined as a deterministic function \( \mu_{\theta}(s) \), which maps each state \( s \) directly to a specific action \( a \). Leveraging the Conditional Expectation Theorem and the deterministic nature of the policy, the objective function \( J(\theta) \) can be expressed as the expected action-value under the state distribution \( \mathcal{S} \):
\[
J(\theta) = \mathbb{E}_{s \sim \mathcal{S}} \left[ Q^{\mu}(s, \mu_{\theta}(s)) \right],
\]
The gradient of this objective with respect to the policy parameters \( \theta \) is given by:
\[
\nabla_{\theta} J(\theta) = \mathbb{E}_{s \sim \mathcal{S}} \left[ \nabla_{\theta} \mu_{\theta}(s) \nabla_{a} Q^{\mu}(a,s) \big|_{a = \mu_{\theta}(s)} \right].
\]
This gradient expression indicates that the policy can be improved by adjusting \( \theta \) in the direction that maximizes the action-value function \( Q^{\mu}(a,s) \).

In practice, the true action value function \( Q^{\mu}(s, a) \) is unknown or computationally expensive to compute directly. Therefore, DPG employs a parameterized approximation \( Q_{\phi}^{\mu}(a,s) \), known as the \textit{critic}, which is trained to approximate \( Q^{\mu}(s, a) \). The objective function then becomes:
\[
J(\theta) = \mathbb{E}_{s \sim \mathcal{S}} \left[ Q_{\phi}^{\mu}(\mu_{\theta}(s), s) \right].
\]
In this context, the policy \( \mu_{\theta} \), parameterized by \( \theta \), is known as the \textit{actor}, since it determines the actions to be taken. 

 The algorithm alternates between updating the critic, \( Q_{\phi}^{\mu} \), and the actor, \( \mu_{\theta} \) as follows:

$\bullet$ {\bf Critic Update}: 
The critic's parameters \( \phi \) ) and consequently the critic, \( Q_{\phi}^{\mu}(a,s) \),  are updated  using temporal-difference learning by minimizing the mean-squared Bellman error:
\[
L(\phi) = {\displaystyle  \mathbb{E}_{(s, a, r, s') \sim \mathcal{D}} }\left[ \left( Q_{\phi}^{\mu}(a,s) -  r - \gamma Q_{\phi}^{\mu'}(\mu'_{\theta}(s'), s')  \right)^2 \right],
\]
where: i) \( s \) and \( s' \) are the current and next states, respectively, ii) \( a \) is the action taken in state \( s \), iii) \( r \) is the reward received for taking action \( a \), iv) \( \gamma \) is the discount factor, and v) \( \mu'_{\theta}(s') \) is the policy determining the action in state \( s' \).

$\bullet$ {\bf Policy Update}: The actor (policy) is updated to maximize the expected return. This is achieved by computing the gradient of the objective \( J(\theta) \) as:
\[
\nabla_{\theta} J(\theta) = \mathbb{E}_{s \sim \mathcal{S}} \left[ \nabla_{\theta} \mu_{\theta}(s) \nabla_{a} Q_{\phi}^{\mu}(a,s) \big|_{a = \mu_{\theta}(s)} \right].
\]
where: i)   \( \mathcal{S} \) is the state space from which \( s \) is sampled, ii) \( \nabla_{\theta} \mu_{\theta}(s) \)is  the gradient of the policy with respect to its parameters, and iii) \( \nabla_{a} Q_{\phi}^{\mu}(s, a) \) is the gradient of the action-value function with respect to the action.

This coupling between the actor and critic allows the actor to improve based on the value estimates from the critic

This framework, where the actor and critic are learned simultaneously, is commonly called an actor-critic method.

Deep Deterministic Policy Gradient (DDPG), introduced by Lillicrap et al.~(2015), extends DPG by using deep neural networks to approximate both the \( Q \)-function \( Q_{\phi}^{\mu}(a,s) \) and the policy \( \mu_{\theta}(s) \). To stabilize learning, DDPG incorporates techniques from Deep Q-Networks (DQN), including replay memory and target networks \cite{mnih2015human}.

\subsubsection*{Multi-Agent Deep Deterministic Policy Gradient (MADDPG)}
The Multi-Agent Deep Deterministic Policy Gradient (MADDPG) algorithm \cite{lowe2017multi} extends the DDPG algorithm to multi-agent environments, supporting both cooperative and competitive tasks. Notably, MADDPG does not require specific assumptions about the environment or the communication channels between agents, such as differentiability. To ensure efficient implementation and scalability, the algorithm is designed so that each agent's policy (actor) relies solely on its own local state information. This design enables a centralized training and decentralized execution framework: during training, information about the full state of all agents is available and used to learn a centralized action-value function (critic), providing a more accurate and comprehensive understanding of the environment and  learning of each agent's policy. Since the learned policies rely only on local state information during execution, agents can act independently without requiring access to global or even neighbors’ information in real time. This makes the approach both practical and efficient for multi-agent systems.

In particular,  let \( N \) denote the number of agents with state vector $\mathbf{s}=\left(s_{1}, \ldots, s_{N}\right)
$, action vector $\mathbf{a}=\left(a_{1}, \ldots, a_{N}\right)$, policy vector  $\boldsymbol{\mu}=\left(\mu_{1}, \ldots, \mu_{N}\right)$,  where each policy \( \mu_{i} \) is parameterized by \( \theta_{i} \). 
Denoting by 
$\left(Q_{\phi_1}^{\mu}, \ldots, Q_{\phi_N}^{\mu}\right)$ the action-value functions parameterized by $\left(\phi_{1}, \ldots, \phi_{N}\right)$, 
the policy for agent $i$ is learned using following gradient update:
\begin{equation}
    \resizebox{0.91\hsize}{!}{
    $\nabla_{\theta_{i}} J(\theta_{i})=\\ \underset{\mathbf{s} \sim \mathcal{S}^N}{\mathbb{E}}\left [\left.\nabla_{\theta_{i}} \mu_{i}( s_{i}) \nabla_{a_{i}} Q_{\phi_i}^{\mu}(\mathbf{a}, \mathbf{s})\right|_{a_{i}=\mu_{i}(s_{i})}\right]$ \nonumber
    }
\end{equation}

The key difference from vanilla DDPG gradient expression is that state and actions of all agents are used to estimate the action-value and are assumed to be accessible at training time. At the same time, each agent has its own action-value function, which allows them to have differing objectives, e.g., competition. 

The critic is learned by minimizing the following loss function:
\begin{equation}
    L\left(\phi_{i}\right)=\underset{\mathbf{s}, \mathbf{a}, \mathbf{r}, \mathbf{s}^{\prime} \sim \mathcal{D}}{\mathbb{E}}\left[\left(Q_{\phi_i}^{\boldsymbol{\mu}}\left(\mathbf{a}, \mathbf{s}\right)-y\right)^{2}\right], 
    \nonumber
\end{equation}
with
$$ y=r_{i}+\left.\gamma Q_{\phi_i}^{\boldsymbol{\mu}^{\prime}}\left( \mathbf{a}^{\prime}, \mathbf{s}^{\prime} \right)\right|_{a_{i}^{\prime}=\mu_{i}^{\prime}\left(s_{j}^{\prime}\right)}.$$

\subsection{MARL in Communication Networks}
\label{sec:sota_MARL_COMM}

Within the framework of modeling dynamic network control problems as a Markov Decision Process (MDP), the shift toward viewing communication networks as a collection of agents with both shared and individual objectives has created new opportunities for applying advanced reinforcement learning techniques. A notable approach is the Multi-Agent Actor-Critic for Mixed Cooperative-Competitive Environments \cite{lowe2017multi}, an extension of the Deep Deterministic Policy Gradient (DDPG) \cite{lillicrap2015continuous} algorithm tailored for multi-agent scenarios. This framework, referred to as MADDPG (Multi-Agent Deep Deterministic Policy Gradient), is highly versatile, supporting both cooperative and competitive tasks without requiring specific assumptions about the network environment.

MADDPG enables agents to learn dynamic policies directly within a simulated network environment, collectively working to maximize a shared reward function. The approach consists of two key phases: during the decentralized execution (inference) phase, each agent acts based on local state observations in the simulated environment, following its own policy function (actor). In the centralized training phase, information from all agents is aggregated to refine the action-value function (critic), ensuring optimal learning outcomes.

Building on this foundation, recent research \cite{wang2024multi} introduced MAFS (Multi-Agent Fine-grained traffic Scheduling), a MADDPG-based packet scheduling system specifically designed for data center networks. Aimed at maximizing throughput while minimizing delay, MAFS incorporates a variant of MADDPG known as CTN-MADDPG (Convolutional Temporal Network -MADDPG). Results demonstrate that MAFS significantly outperforms traditional MADDPG, DDPG, and ECMP (Equal-Cost Multi-Path) baseline approaches, achieving superior high-throughput, low-latency scheduling.

Another notable approach is the Recurrent Softmax Delayed Deep Double Deterministic policy gradient (RSD4) algorithm, introduced in \cite{hu2022effective}. RSD4 addresses delay-constrained scheduling as a Partially Observable Markov Decision Process (POMDP), leveraging a Lagrange dual update mechanism. While the algorithm demonstrates significant improvements over traditional deterministic scheduling methods like Earliest Deadline First (EDF) in managing delay constraints within dynamic network environments, its applicability to real-world scenarios is limited by several factors.
First, the network topologies considered in RSD4 are highly simplified, with a maximum of three paths per network, each supporting only a single flow. Furthermore, these paths often lack intersections or shared communication interfaces, making the results less representative of the complexities found in practical deployments. Additionally, RSD4 employs a Recurrent Neural Network (RNN) architecture for scheduling, which, while powerful, introduces unnecessary computational overhead compared to simpler and more efficient alternatives like Multi-Layer Perceptrons (MLPs).
Finally, the algorithm’s scope is restricted to training a scheduler, with no mechanism for dynamic routing or adapting to varying traffic patterns. This limits its utility in environments where routing decisions play a critical role in optimizing network performance.

In \cite{liu2023deadline}, the authors propose the "Deadline-constrained Multi-agent Collaborative Transmission" framework, a fully distributed solution for reliable transmission in dynamic networks with delay-sensitive applications.  The framework leverages In-band Network Telemetry (INT) for real-time network status detection and formulates a joint optimization problem that considers flow deadlines, queuing delays, and network conditions. 
To solve this problem, the authors employ the Asynchronous Advantage Actor-Critic (A3C) algorithm \cite{mnih2016asynchronous}, a training method originally developed for single-agent systems but adapted to handle multi-agent scenarios.
Despite its innovative approach, DMCT has several limitations. While the framework is designed for fully distributed execution, it relies on a non-machine learning algorithm for centralized action computation. Specifically, a centralized agent calculates the actions for distributed routers and periodically sends updates, which can limit scalability and real-time adaptability. Furthermore, the method does not enforce strict deadlines; packets with excessive accumulated delay are simply dropped once their deadlines are exceeded. Additionally, while A3C is used for training, its original design for single-agent systems makes its adaptation to multi-agent scenarios less robust. Finally, the practical implementation of DMCT heavily depends on the authors' proprietary technology, which may restrict its applicability to broader use cases.

\section{System Model}
\label{sec:system_model}

Given the nature of modern time-critical applications, it is necessary to define specific network, service, and queuing models that are capable of expressing the requirements of such applications.

\subsection{Network and service parameters}
\label{sec:network_and_service_parameters}

We consider a communications network described by a directed graph \( \mathcal{G} = (\mathcal{V}, \mathcal{E}) \), where \( \mathcal{V} \) and \( \mathcal{E} \) denote the set of nodes and links, respectively. 

We use $\rho_{i}^{+} \subset \mathcal{V}$  and $\rho_{i}^{-} \subset \mathcal{V}$ to denote the set of outgoing and incoming neighbors of node $i \in \mathcal{V}$, respectively. 

The system operates in time-slotted intervals with equally-sized slots, indexed by \( t \in \{0, 1, \ldots \} \). For each link \( (i, j) \in \mathcal{E} \), the link capacity \( C_{ij}(t) \) specifies the number of packets that can be transmitted over link \( (i, j) \) during time slot \( t \). 

The network supports latency-sensitive services that require the timely delivery of packets across multiple source-destination pairs. Quality of Service (QoS) requirements are imposed by associating a maximum {\em lifetime}, or Time to Live (TTL), to each service packet. A packet with a positive lifetime is considered effective and continues to flow through the network. Conversely, a packet is considered outdated once its lifetime reaches zero and is immediately dropped from the network.

We identify time-sensitive services as a set of commodities, where each commodity \( c \in \mathcal{C} \) is associated with:
\begin{itemize}
    \item  a source node \( s^{c} \in \mathcal{V} \),
    \item a destination node \( d^{c} \in \mathcal{V} \), and
    \item an initial lifetime \( L^c \in \mathcal{L} = \{1, \dots, L_{\text{max}}\} \).
\end{itemize}
Note that a lower initial lifetime $L^{c}$ indicates a more latency-sensitive service. 
The stochastic number of commodity-$c$ packets arriving at source node $s^{c}$  at time $t$ is denoted by $b^{c}(t)$, with $\bar{b}^{c} = \mathbb{E} \big[ b^{c}(t) \big]$ denoting its mean arrival rate. We use $\bm{b}(t) \triangleq \big\{b^{c}(t), \forall c \in \mathcal{C}\big\}$ to denote the packet arrival vector at time $t$.

\subsection{Network control variables}

We consider network control policies that perform packet routing and scheduling decisions, represented by the flow variables \( f_{ij}^{(c, \ell)}(t) \). Here, \( f_{ij}^{(c, \ell)}(t) \) denotes the number of packets of commodity \( c \in \mathcal{C} \) with lifetime \( \ell \) transmitted over link \( (i, j) \in \mathcal{E} \) at time \( t \).

Furthermore, 
${\bf f}(t) \triangleq \big\{ f_{ij}^{(c,\ell )}(t), \forall (i,j) \in \mathcal{E}, \forall c\in\mathcal{C}, \ell \in \mathcal{L} \big\}$ 
denotes the entire set of flows across the network at time \( t \)

For ease of notation, we also use 
\[
f_{i \rightarrow}^{(c, \ell)}(t) \triangleq \sum_{j \in \rho_{i}^{+}} f_{ij}^{(c, \ell)}(t),
\quad
f_{\rightarrow i}^{(c, \ell)}(t) \triangleq \sum_{j \in \rho_{i}^{-}} f_{ji}^{(c, \ell)}(t),
\]

to denote the total outgoing flow and the total incoming flow from/to node $i$, respectively. 

To account for policies that actively drop packets from the network before their lifetime expires, we introduce the packet-dropping variables \( g_{i}^{(c, \ell)}(t) \). These variables represent the number of packets belonging to commodity \( c \) with lifetime \( \ell \) that are intentionally dropped by node \( i \) during time slot \( t \). 

The complete set of packet-dropping variables across the network at time \( t \) is defined as:
\[
\mathbf{g}(t) \triangleq \big\{ g_{i}^{(c, \ell)}(t) \mid i \in \mathcal{V}, c \in \mathcal{C}, \ell \in \mathcal{L} \big\}.
\]

This formulation allows for the explicit modeling of proactive packet-dropping policies, which can help optimize network performance by balancing delay constraints and resource availability.

\subsection{Lifetime Queue Dynamics}

Upon arrival, packets are added to network queues based on corresponding lifetimes and the commodity they belong to. The queue backlog of lifetime $\ell$ packets at node $i$ for commodity $c$ on time slot $t$ is denoted as $q_{i}^{(c,\ell)}(t)$, and ${\bf q}(t) \triangleq \{ q_{i}^{(c,\ell)}(t), \forall i \in \mathcal{V}, \ell \in \mathcal{L}, c \in \mathcal{C} \}$ denotes the queuing state vector. 
Queue dynamics for the lifetime packets evolve as: 
\begin{align}
    & \blue{q_{i}^{(c, \ell)}(t) =  q_{i}^{(c,\ell + 1)}(t-1) \! - \! f_{i \rightarrow}^{(c,\ell + 1)}(t-1) - g_{i}^{(c,\ell+1)}(t-1)} \notag\\
    & \blue{+  f_{\rightarrow i}^{(c,\ell + 1)}(t-1) + b_{i}^{(c,\ell)}(t),} 
    \label{eq:lifetime-queue}
\end{align}

$\forall i \in \mathcal{V}, \forall l \in \mathcal{L},\forall c \in \mathcal{C}, \forall t$  with:
\begin{equation*}
  b_{i}^{(c,\ell)}(t) =
    \begin{cases}
      b^{c}(t) & \text{if $i = s^{c}, \ell = L^{c}$}\\
      0 & \text{otherwise}
    \end{cases}       
\end{equation*}

We assume that packets with $0$ lifetime, namely {\em expired packets}, are immediately dropped from the queue backlog:
\begin{align} \label{eq:expired-packets}
    q_{i}^{(c,0)}(t) &= 0, &&\forall i \in \mathcal{V}, \forall c \in \mathcal{C} , \forall t
\end{align}
While packets reaching to the destination will be immediately consumed:
\begin{align}
    q_{d_{c}}^{(c,\ell)}(t) &= 0, && d_{c} \in \mathcal{V}, \forall \ell \in \mathcal{L}, \forall c \in \mathcal{C}, \forall t
    \label{eq:packets-destination}
\end{align}


\section{DCMT Problem Formulation}
\label{sec:DCMT_Problem_Formulation}

The objective of the Delay-Constrained Maximum-Throughput (DCMT) network control problem consists of maximizing the total number of packets delivered within their respective deadlines, referred to as aggregate timely throughput: 
\begin{equation}
    \lim_{T \to \infty} \frac{1}{T} \sum_{t=0}^{T-1} \sum_{c \in \mathcal{C}} \sum_{\ell \in \mathcal{L}} \mathbb{E}\left[f^{(c, \ell)}_{\rightarrow d_c}(t)\right]
\end{equation}

\noindent where $f^{(c, \ell)}_{\rightarrow d_c}(t)$ denotes the flow of packets delivered to destination $d_c$ for commodity $c$ with lifetime $\ell$ at time $t$.

\subsection{Availability Constraint}
The total outgoing flow (e.g., forwarded and dropped packets) for each commodity must not exceed its current queue backlog at each node:
\begin{equation}
    f^{(c, \ell)}_{i \rightarrow}(t) + g^{(c, \ell)}_i(t) \leq q^{(c, \ell)}_i(t), \quad \forall i \in \mathcal{V}, c \in \mathcal{C}, \ell \in \mathcal{L}, \forall t
\end{equation}

\subsection{Capacity Constraint}
To prevent overflow and maintain feasible network flows, the flow over each link is bounded by its capacity:

\begin{equation}
    \sum_{c \in \mathcal{C}} \sum_{\ell \in \mathcal{L}} f^{(c, \ell)}_{ij}(t) \leq C_{ij}(t), \quad \forall (i, j) \in \mathcal{E}, c \in \mathcal{C}, \ell \in \mathcal{L}, \forall t
\end{equation}

\subsection{Problem Formulation}
Therefore, the goal of the DCMT problem is to maximize the aggregate timely throughput subject to queue dynamics, availability, and capacity constraints, i,e.: 
\begin{align} 
&\text{max} && \lim_{T \rightarrow \infty} \frac{1}{T} \sum_{t = 0}^{T-1} \sum_{c\in\mathcal{C}} \sum_{\ell \in \mathcal{L}} \mathbb{E}[f_{\rightarrow d_{c}}^{(c,\ell)}(t)] \label{eq:timely_thr_max}  \\
&\text{s.t.}  && \text{Queues evolve according to \eqref{eq:lifetime-queue}-\eqref{eq:packets-destination}} && \nonumber \\
 &&& \hspace{-0.75cm}f_{i\rightarrow}^{(c,\ell)}\!(t) \!+\! g_{i}^{(c,\ell)}\!(t) \!\leq\! q_{i}^{(c,\ell)}\!(t) && \hspace{-1.35cm} \forall i \in \mathcal{V}, \forall c \in \mathcal{C}, \forall \ell \in \mathcal{L}, \forall t \nonumber \\
     &&&\hspace{-0.75cm} \sum_{\ell \in \mathcal{L}} \sum_{c \in \mathcal{C}} f_{ij}^{(c,\ell)}(t) \leq C_{ij}(t)  && \hspace{-0.75cm} \forall (i,j) \in \mathcal{E}, \forall c \in \mathcal{C}, \forall t \nonumber \\
     &&&\hspace{-0.75cm} f_{ij}^{(c,\ell)}(t) \geq 0 && \hspace{-1.7cm}\forall (i,j) \in \mathcal{E}, \forall c \in \mathcal{C}, \ell\in\mathcal{L} ,\forall t \nonumber 
\end{align}

While the above network control problem cannot be reduced to a stochastic optimization problem~\cite{neely2022stochastic, cai2022ultra}, it can be modeled as an MDP and effectively addressed via RL.

\section{MADRL Network Control Framework} 
\label{sec:framework}

The proposed Multi-Agent Deep Reinforcement Learning (MADRL) framework adopts a centralized routing and distributed scheduling architecture, as outlined in Section \ref{sec:arch_description}. This framework integrates the Multi-Agent Actor-Critic paradigm, specifically leveraging the Multi-Agent Deep Deterministic Policy Gradient (MADDPG) method, combined with a design process that incrementally fosters a deeper synergy between data-driven and model-driven approaches, enabling  a deeper understanding of the key factors impacting network performance.
By doing so, it equips agents with the ability to dynamically balance network throughput and latency requirements. Through learning optimal routing and scheduling strategies, the framework effectively addresses the complexities and constraints outlined in the DCMT network control problem, providing a scalable and adaptive solution for managing delay-sensitive applications in dynamic network environments.

\subsection{Centralized Routing and Distributed Scheduling}
\label{sec:arch_description}

In the proposed centralized routing and distributed scheduling architecture, a single centralized routing agent dynamically assigns paths to newly arrived packets based on network congestion, ensuring efficient utilization of network resources. Simultaneously, multiple fully distributed scheduling agents make local decisions to prioritize packets according to their residual lifetimes and assigned paths. 
Working in tandem, these agents collaboratively learn cooperative policies designed to maximize aggregate timely throughput, as outlined in Section \ref{sec:DCMT_Problem_Formulation}.
 
Each agent, $x$, (router or scheduler) operates according to its policy, $\mu_{x}(s_{x}(t)) = a_{x}(t)$, which based on the agent's observed state $s_{x} \in \mathcal{S}$ at time $t$, selects an action $a_{x} \in \mathcal{A}$ which influences the communication network described in Section \ref{sec:system_model}. Agents are trained collaboratively based on the actor-critic algorithm, whereby each agent's actor-network (policy) is optimized through a global critic (value function) with complete visibility of agents' actions and states. This training process aims to let the agents develop an emergent cooperative behaviour based solely on local state.
This collaborative design leverages both local and global insights to adaptively control the network in dynamic environments, effectively addressing the limitations of traditional stochastic optimization methods when faced with high congestion and strict delay requirements.

\subsection{MADRL Framework network and service model}
We expand the network and service model from Section \ref{sec:network_and_service_parameters} to incorporate additional parameters introduced by using a centralized router and distributed schedulers\footnote{\blue{For ease of presentation, throughout the paper, we use routing agent and router, and scheduling agent and scheduler, interchangeably}.} that rely on the assignment of packets to specific paths upon their arrival to their respective sources. 

Let $\mathcal{P}^c$ denote the set of candidate paths for each commodity $c \in \mathcal{C}$, which consists of all feasible\footnote{\blue{ A path is said to be feasible for commodity~$c$ if its length does not exceed 
the initial lifetime of commodity~$c$. Formally, the set of feasible paths for commodity~$c$ consists of all source--destination paths whose traversal time, in the absence of queuing delays, is within the initial lifetime of commodity~$c$. This set depends only on the network topology and the commodity's attributes (initial lifetime and source--destination pairs)  and is computed offline.}} paths from the source to the destination of commodity $c$. The set of all commodity paths is denoted as  $\mathcal{P}=\bigcup_{c \in \mathcal{C}} \mathcal{P}^c$. Furthermore, we define $\mathcal{P}_{ij}^{c} = \{ p \in \mathcal{P}^c \mid (i, j) \in p \}$ as the subset of paths for commodity $c$ that traverse link $(i, j)$.

Accordingly, network queues are now designed to accumulate packets also based on their assigned path. Hence, the queue backlog at node $i$ of packets of commodity $c$, lifetime $\ell$, and path  $p$, at time $t$ is denoted by $q_{i}^{(c, p,\ell)}(t)$,

and ${\bf q}(t) \triangleq \{ q_{i}^{(c, p,\ell)}(t), \forall i \in \mathcal{V}, c \in \mathcal{C}, p \in \mathcal{P}^c, \ell \in 1 \ldots L^c \}$ denotes the network's queuing state vector. \blue{We also use ${\bf q_{i}}(t) \triangleq \{ q_{i}^{(c, p,\ell)}(t), c\in \mathcal{C}, p \in \mathcal{P}^c, \ell \in 1 \ldots L^c \}$ to denote the queuing state of node $i\in\mathcal V$, and ${\bf q_{ij}}(t) \triangleq \{ q_{i}^{(c, p,\ell)}(t), c\in \mathcal{C}, p \in \mathcal{P}^c_{ij}, \ell \in 1 \ldots L^c \}$ to denote the queuing state of interface $(i,j)\in\mathcal E$}. 
\blue{Analogously, we denote the congestion state of node $i$ and interface $(i,j)$ as $Q_{i}(t) = \sum_{q \in q_{i}(t)} |q|$ and $Q_{ij}(t) = \sum_{q \in q_{ij}(t)} |q|$, respectively. The congestion state of link or interface $(i,j)$ refers to the number of packets at node $i$ assigned to any path that uses link $(i,j)$. That is, those packets waiting at node $i$ to traverse link $(i,j)$.}

Finally, we use $f_{ij}^{(c,p,\ell)}(t)$ to denote the number of packets of commodity $c$, lifetime $\ell$, and path $p$, flowing over link $(i,j)\in \mathcal{E}$ at time $t$, and ${\bf f}(t) \triangleq \big\{ f_{ij}^{(c, p,\ell )}(t), \forall (i,j) \in \mathcal{E}, \forall c \in \mathcal{C}, \forall p\in\mathcal{P}, \forall \ell \in \mathcal{L} \big\}$ to denote the network flow vector. 
In addition, the total outgoing and incoming flow from/to node $i$ are given by $f_{i \rightarrow}^{(c,p,\ell)}(t) \triangleq \sum_{j \in \rho_{i}^{+}} f_{ij}^{(c, p,\ell)}(t)$ and $f_{\rightarrow i}^{(c,p,\ell)}(t) \triangleq \sum_{j \in \rho_{i}^{-}} f_{ji}^{(c,p,\ell)}(t)$.

\begin{figure}
    \centering
    \includegraphics[width=0.5\textwidth]{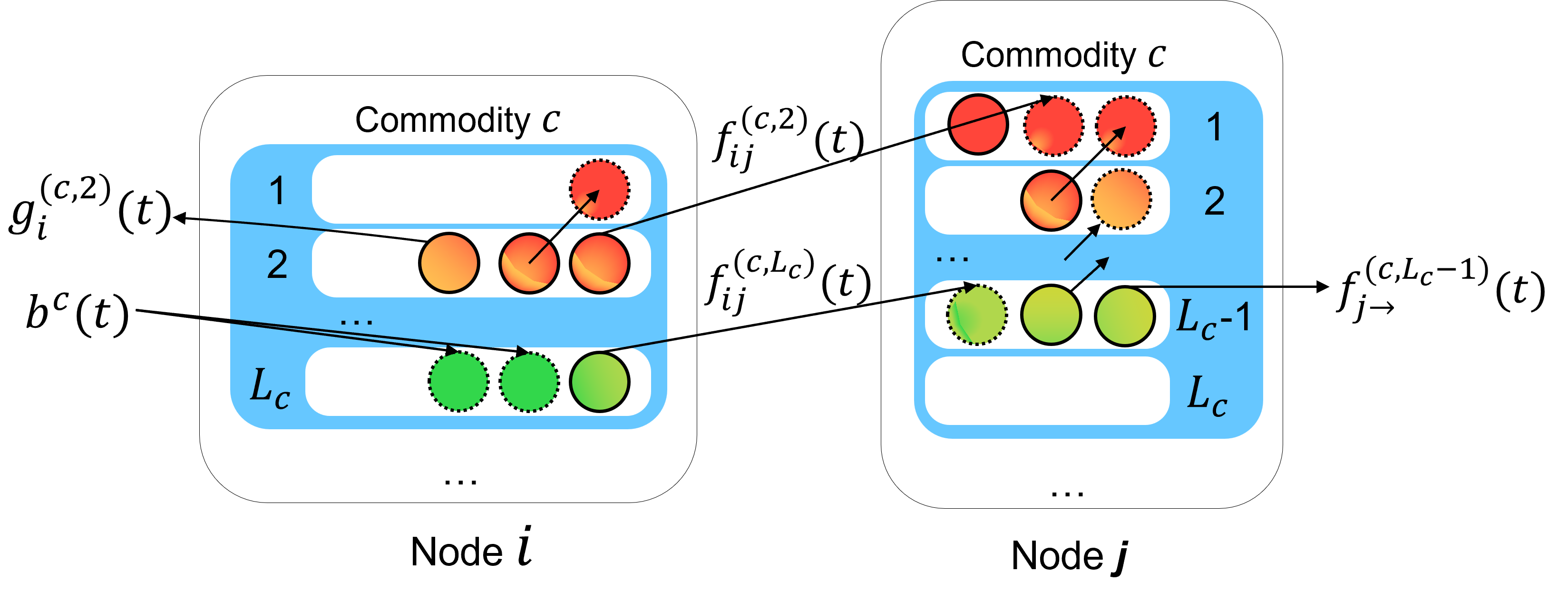}
    \caption{\small{Queue dynamics illustration. Packet colors from green to red indicate higher to lower lifetimes. Solid and dashed packets denote current and next time slot, respectively.}}
    \label{fig:unina_Queue}
    \vspace{-0.45cm}
\end{figure}

\section{MADRL Network Control Strategies}
\label{sec:network_control_strategies}

This section outlines the strategies developed within the proposed centralized routing and distributed scheduling architecture described in section~\ref{sec:arch_description}. 
All strategies share the same routing agent and differ on the scheduling agents' design, focusing on two key aspects: the size of the agents' state and action spaces, and the agents' inference time complexity.
Note that, for each RL agent, the \textbf{inference time complexity} is given by two components. The first one is the time needed to generate the action $a(t)$ based on the state $s(t)$, which depends on the neural network's architecture, the size of the state space, and the size of the action space. The second component is driven by the algorithm that applies the action into the network. 
We assume that every RL agent's Neural Network (NN) architecture is a Multi-Layer Perceptron (MLP) with two hidden layers with respectively $m_{1}$ and $m_{2}$ hidden neurons. Choosing an MLP-based architecture ensures that the NN remains computationally efficient while maintaining sufficient expressive power to address network control problems with RL-based approaches. It also provides a direct mapping between the state and action spaces without any preprocessing or intermediate operations. In section \ref{sec:strategy_nn_improvements}, we discuss further aspects regarding the proposed strategies and the choice of the NN architecture.

The following subsections present the MADRL-based strategies proposed in this paper. Each subsection introduces an incrementally enhanced scheduling strategy, focusing on key aspects such as optimized action space design and efficient state space representation. 
Building on prior strategies, each approach deepens the integration of data-driven and model-driven methodologies.
In Section \ref{sec:base_maddpg}, we describe the initial and most intuitive MADRL strategy, where the network naturally drops packets upon lifetime expiry, and the schedulers account for three decisions, namely how many packets (proactively) drop, send and keep. 
Then, in  Section \ref{sec:no_drop}, we improve the base strategy by relieving the schedulers' from the burden of proactively dropping packets, leaving expiration by aging as the only way to drop packets from the system. Section \ref{sec:effective_lifetime} introduces the \textit{effective lifetime}, a more efficient way of representing the packets' residual lifetime that impacts the size of both the action and the state spaces, and reduces the number of packets flowing through the network without any chance of reaching the destination. 
Further observing that under centralized routing, there is no benefit of holding packets if there is available capacity in their corresponding outgoing interface, the strategy in Section \ref{sec:send_max}, further reduces the schedulers' actions by letting them only decide which packets to choose to fill up the outgoing interface capacity.

Building on insights from the previous strategies, Section \ref{sec:lelf_heuristic} introduces a MARL strategy where schedulers follow a predetermined policy. Specifically, this approach integrates a centralized MLP-based router with a distributed rule-based scheduling policy, LELF, which prioritizes packets with the lowest effective lifetime.
Finally, in \ref{sec:mec_lelf}, we test a full rule-based strategy that combines a minimum link-weight router with distributed LELF schedulers.

\subsection{MARL Base Strategy: Lifetime Drop/Send/Keep}
\label{sec:base_maddpg}
We first describe the centralized \textbf{Routing Agent}, common to all proposed policies. 
The state and action spaces of the routing agent are defined as follows: 
\begin{itemize}
    \item The \textbf{State} space has size $|\mathcal{C}|+|\mathcal{V}|$ and consists of the exogenous arrival of new packets in the network combined with the congestion state of nodes, $s_{router}(t)=\{ b^{c}(t),.., Q_{i}(t), \forall c\in\mathcal{C}, \forall i \in \mathcal{V} \} $.
    \item The \textbf{Action} space has size $|\mathcal{P}|$,  and represents the assignment of newly arrived packets to each path, denoted by $a_{router}(t)=\{ \mathcal{M}^{(p)}, \forall c\in \mathcal{C}, p\in P^c\}$, where  $\mathcal{M}^{(p)}$ is the number of packets assigned to path $p$.
\end{itemize}

For this initial strategy, the state and action spaces of the \textbf{Scheduling Agent} managing interface $(i,j)$ are defined as follows.
\begin{itemize}
    \item The \textbf{State} space has size $\sum_{c\in\mathcal{C}} (L^c*|P_{ij}^{c}|)$ and consists of the queuing state vector \blue{of interface 
    $(i,j)$, $s_{sch_{ij}}(t) = {\bf q_{ij}}(t)$}. 
    \item \blue{The \textbf{Action} space has size $
\sum_{c \in \mathcal{C}} \bigl( 3 \cdot L^c \cdot |P_{ij}^c| \bigr)$,
and consists, for each commodity $c \in \mathcal{C}$, path $p \in P_{ij}^c$, and lifetime $\ell \in \{1,\ldots,L^c\}$, of a triplet
$a^{(c,p,\ell)}_{sch_{ij}}(t) = \bigl[g^{(c,p,\ell)},\, F^{(c,p,\ell)},\, K^{(c,p,\ell)}\bigr]$, 
where $g^{(c,p,\ell)}$, $F^{(c,p,\ell)}$, and $K^{(c,p,\ell)}$ denote, respectively, the number of packets to \emph{drop}, \emph{forward}, and \emph{keep} for commodity $c$, path $p$, and lifetime $\ell$ at node $i$.}
\end{itemize}
The way in which the scheduling agent's action is applied to the network is reported in Algorithm \ref{alg:marl_scheduler_dsk}.
\begin{algorithm}
	\caption{MARL LT D/S/K -- Scheduler}
    \begin{algorithmic}[1]
        \For \blue{{$c \in \mathcal{C}$}}
            \For {$p \in P_{ij}^{c}$}
                \For{$\ell \in \{ 1 \ldots L^c$\}}
                    \State \blue{$[g, F, K ]= a_{sch_{ij}}^{(c,p,\ell)}(t) $}
                    \State $q_{i}^{(c, p,\ell)} \leftarrow q_{i}^{(c, p,\ell)} -g -F$
                    \State $f_{ij}^{(c, p,\ell)} \leftarrow f_{ij}^{(c, p,\ell)} +F$
                \EndFor
			\EndFor
        \EndFor
	\end{algorithmic}
    \label{alg:marl_scheduler_dsk}
\end{algorithm}

The time complexity for the routing agent is given by the combination of: 
\begin{itemize}
    \item The time to generate action with the MLP NN is $\bigO((|\mathcal{C}|+|\mathcal{V}|)*m_{1}+m_{1}*m_{2}+m_{2}*|\mathcal{P}|)$
    \item The time to apply the action is $\bigO(|\mathcal{P}|)$.
\end{itemize}
The time complexity for the scheduling agent is given by:
\begin{itemize}
    \item The time to generate the action with the MLP NN is $\bigO((\sum_{c\in\mathcal{C}} (L^c*|P_{ij}^{c}|))*m_{1}+m_{1}*m_{2}+m_{2}*(\sum_{c\in\mathbf{C}} (3*L^c*|P_{ij}^{c}|)))$ 
    \item The time to apply the action,  based on Algorithm \ref{alg:marl_scheduler_dsk}, is $\bigO(|\mathcal{P}|*|\mathcal{L}|*3)$.
\end{itemize}

\subsection{MARL Strategy Improvement 1: Lifetime Send/Keep} 
\label{sec:no_drop}

This strategy refines the Lifetime Drop/Send/Keep approach from Section \ref{sec:base_maddpg}. The {\bf Router} remains unchanged, but the {\bf Schedulers}' action space is reduced by removing the drop action, streamlining training with fewer action choices.
The intuition behind this improvement is to leverage the natural evolution of packets lifetime, which reduces until expiration as time flows, in order to relieve agents from the burden of deciding when it is more convenient to proactively drop a packet rather than to forward or keep it. As a consequence, the {\bf Schedulers}' action space is refined as follows:

\begin{itemize}
    \item The \textbf{Action} space size reduces to $\sum_{c\in\mathbf{C}} (2*L^c*|P_{ij}^{c}|)$ and consists, \blue{ for each commodity $c \in \mathcal{C}$, path $p \in P_{ij}^c$, and lifetime $\ell \in \{1,\ldots,L^c\}$, of a couple $ a^{(c,p,\ell)}_{sch_{ij}}(t) = \bigl[ F^{(c,p,\ell)},\, K^{(c,p,\ell)}\bigr]$ denoting the number of packets to \emph{forward} and \emph{keep}.} The Algorithm \ref{alg:marl_scheduler_dsk} is adapted as described in Algorithm \ref{alg:marl_scheduler_sk}.
\end{itemize}

As a consequence, the time complexity becomes:

\begin{itemize}
    \item The time to generate the action with the MLP NN is $\bigO((\sum_{c\in\mathcal{C}} (L^c*|P_{ij}^{c}|))*m_{1}+m_{1}*m_{2}+m_{2}*(\sum_{c\in\mathbf{C}} (2*L^c*|P_{ij}^{c}|)))$.
    \item The time to apply the action, based on Algorithm \ref{alg:marl_scheduler_sk}, is $\bigO(|\mathcal{P}|*|\mathcal{L}|*2)$.
\end{itemize}

\begin{algorithm}
	\caption{MARL LT S/K -- Scheduler}
	\begin{algorithmic}[1]
        \For {\blue{$c \in \mathcal{C}$}}
            \For {$p \in P_{ij}^{c}$}
                \For{$\ell \in \{ 1 \ldots L^c$\}}
                    \State \blue{$[F, K]= a_{sch_{ij}}^{(c,p,\ell)}(t) $}
                    \State $q_{i}^{(c, p,\ell)} \leftarrow q_{i}^{(c, p,\ell)} -F$
                    \State $f_{ij}^{(c, p,\ell)} \leftarrow f_{ij}^{(c, p,\ell)} +F$
                \EndFor
			\EndFor
        \EndFor
	\end{algorithmic}
    \label{alg:marl_scheduler_sk}
\end{algorithm}

Although moderate, this first improvement already yields a reduction of the Scheduler's inference time complexity. In addition, we should expect a faster training convergence.

\subsection{MARL Strategy Improvement 2: Effective Lifetime Send/Keep}
\label{sec:effective_lifetime}
 This strategy builds on Lifetime Send/Keep from Section \ref{sec:no_drop}, \blue{improving the way routing information can be leveraged to more accurately determine the usefulness of keeping a packet in the network. To this end, we introduce the concept of \textit{{\bf Effective Lifetime}}, which denotes the maximum number of time slots a packet can wait in a network queue before having no chance to reach its destination over its assigned path. This is computed as the difference between a packet's lifetime (time until a packet becomes useless to the application) and the minimum time required to reach its destination over its assigned path (i.e., under our assumption of one timeslot per network hop, the length of the remaining path until the destination).
 Mathematically, we define the effective lifetime of a packet assigned to path $p$, currently at node $i$, with lifetime $\ell$, denoted as $\blue{EL(p,\ell,i)}$, as:}  
 \begin{equation}
    \label{eq:effective_lifetime}
    \blue{\blue{EL(p,\ell,i)}= \ell - dist(i,p) + 1 
    \,\, \forall p \in \mathcal{P}, \forall i \in p}
 \end{equation}
 \blue{With $dist(i,p)$ denoting the distance between node $i$ and the destination of path $p$.}
 Compared to the {\em absolute} lifetime, the {\em effective} lifetime decreases only when packets wait in a queue and remains unchanged when packets traverse a network link.
 \blue{Under this definition, we have that (i) a packet can only reach its destination if it has positive effective lifetime, (ii) the effective lifetime of a packet is always smaller or equal than its absolute lifetime. Hence, making the network drop packets when their effective lifetime goes to zero has important benefits.
 On the one hand, it reduces congestion and queue size, since packets with no chance of arriving at the destination are dropped instead of being retained until absolute lifetime expiration.  
 On the other hand, representing the state of the network under effective lifetime significantly reduces agents' state and action spaces.}

While the {\bf Router} design remains unchanged, the \textbf{Schedulers}' states and actions now account for this enhanced lifetime metric as follows:

 \begin{itemize}
     \item The \textbf{State} space size reduces to 
     $\sum_{c\in\mathbf{C}}\sum_{p\in P_{ij}^c}\blue{EL(p,L^c,i)}$. 
     \item The \textbf{Action} space size reduces to $\sum_{c\in\mathbf{C}}\sum_{p\in P_{ij}^c} (2*\blue{EL(p,L^c,i)})$ with $\blue{EL(p,L^c,i)} \ll L^c, \forall p \in P^c, \forall i \in p$.
 \end{itemize}
 
{The time complexity of a scheduler action changes accordingly:
\begin{itemize}
    \item The time to generate the action with the MLP NN is $\bigO( (\sum_{c\in\mathbf{C}}\sum_{p\in P_{ij}^c}\blue{EL(p,L^c,i)})*m_{1}+m_{1}*m_{2}+m_{2}*(\sum_{c\in\mathbf{C}}\sum_{p\in P_{ij}^c} (2*\blue{EL(p,L^c,i)}))$.
    \item The time to apply the action,  based on Algorithm \ref{alg:marl_el_scheduler_sk}, is $\bigO(|\mathcal{P}|*(\max_{p \in \mathcal{P}} \blue{EL(p,L_{max},i)})*2)$.
\end{itemize}

\begin{algorithm}
	\caption{MARL EL S/K-- Scheduler}
	\begin{algorithmic}[1]
		\For {\blue{$c \in  \mathcal{C}$}}
            \For {$p \in P_{ij}^{c}$}
                \State $el = \blue{EL(p,L^c,i)}$
                \For{$\ell \in \{ 1 \ldots el \} $}
                    \State \blue{$[F, K ]= a_{sch_{ij}}^{(c,p,\ell)}(t) $}
                    \State $q_{i}^{(c, p,\ell)} \leftarrow q_{i}^{(c, p,\ell)} -F$
                    \State $f_{ij}^{(c, p,\ell)} \leftarrow f_{ij}^{(c, p,\ell)} +F$
                \EndFor
			\EndFor
		\EndFor
	\end{algorithmic}
    \label{alg:marl_el_scheduler_sk}
\end{algorithm}

\subsection{MARL Strategy Improvement 3: Effective Lifetime S-Max}
\label{sec:send_max}
Based on the \textit{Effective Lifetime} concept and on \textit{MARL Effective Lifetime Send/Keep} strategy introduced in section \ref{sec:effective_lifetime}, this strategy further reduces the Schedulers' action space by removing the keep action. The resulting scheduler's action is described by a probability distribution $a_{sch_{ij}}$ used to select packets from the node queues to be sent over link $(i,j)$ until the link capacity is reached or until there are no more \emph{eligible} packets to be forwarded.
This is motivated by the fact that in a max-throughput network control problem where routes are assigned and fixed at the source, there is no benefit from keeping packets that could have been sent over their corresponding interfaces.
The new scheduling agent design, therefore, uses the probabilistic function $a_{sch_{ij}}$ to choose commodity-path-lifetime packets to fill up the outgoing link capacity based on Algorithm \ref{alg:marl_el_scheduler_smax}.

A direct consequence of these design choices is a further  reduction of the resulting \textbf{Schedulers}' action space (recall that the \textbf{Router} design remains unchanged): 
\begin{itemize}
    \item The \textbf{Action} space size is further reduced to $\sum_{c\in\mathbf{C}}\sum_{p\in P_{ij}^c} \blue{EL(p,L^c,i)}$ which now consists, \blue{for each commodity $c \in \mathcal{C}$, path $p \in P_{ij}^c$, and lifetime $\ell \in \{1,\ldots,\blue{EL(p,L^c,i)}\}$ of a single component 
    $a^{(c,p,\ell)}_{sch_{ij}}(t) = \bigl[\probP_{F^{(c,p,\ell)}}\bigr]$, which is interpreted as the probability to \emph{forward} the associated packet}.
\end{itemize}

The resulting time complexity for the scheduler is:

\begin{itemize}
    \item The time to generate the action with the MLP NN is $\bigO( (\sum_{c\in\mathbf{C}}\sum_{p\in P_{ij}^c}\blue{EL(p,L^c,i)})*m_{1}+m_{1}*m_{2}+m_{2}*(\sum_{c\in\mathbf{C}}\sum_{p\in P_{ij}^c} \blue{EL(p,L^c,i)}))$.
    \item The time to apply the action,  based on Algorithm \ref{alg:marl_el_scheduler_smax}, is $\bigO(|q_{i}|*|\mathcal{P}|*(\max_{p \in \mathcal{P}} \blue{EL(p,L_{max},i)}))$.
\end{itemize}

\begin{algorithm}
	\caption{MARL EL S-MAX -- Scheduler}
	\begin{algorithmic}[1]
        \State $res\_flow \leftarrow C_{ij}$
        \While {$res\_flow > 0 $ and $eligible(q_{i},a_{sch_{ij}})$}
            \For{\blue{$c \in \mathcal{C}$}}
                \For {$p \in P_{ij}^{c}$}
                    \State $el = \blue{EL(p,L^c,i)}$
                    \For{$\ell \in \{ 1 \ldots el \} $}
                        \State $F = 0$
                        \State \blue{$\probP_F = a_{sch_{ij}}^{(c,p,\ell)}(t)$}
                        \If{$Take(\probP_{F}, q_{i}^{(c, p,\ell)})$}
                            \State $F = q_{i}.deque(c, p,\ell,1)$
                            \State $f_{ij}^{(c, p,\ell)} \leftarrow f_{ij}^{(c, p,\ell)} +F$
                            \State $res\_flow = res\_flow - F$
                        \EndIf
                    \EndFor
    			\EndFor
            \EndFor
        \EndWhile
	\end{algorithmic}
    \label{alg:marl_el_scheduler_smax}
\end{algorithm}

\subsection{MARL Strategy Improvement 4: Effective Lifetime LELF}
\label{sec:lelf_heuristic}

Based on the improvements driven by the introduction of the Effective Lifetime in sections \ref{sec:effective_lifetime} and \ref{sec:send_max}, we introduce a simpler rule-based scheduling policy named Lower Effective Lifetime First (LELF).
In order to maximize the number of packets delivered on time, the LELF scheduling policy favors forwarding packets with the lowest effective lifetime. Recall that as long as packets keep moving, their effective lifetime does not decrease, so the goal is to keep as many packets as possible with a positive effective lifetime, i.e., with a chance of arriving at their destinations. Hence, the LELF policy attempts to fill the link capacity by selecting packets in increasing order of effective lifetime. 

While the \textbf{Router} design remains unchanged, the \textbf{Schedulers}' action space now account for this rule-based policy:

\begin{itemize}
    \item The \textbf{Action} space size remains at $\sum_{c\in\mathbf{C}}\sum_{p\in P_{ij}^c} \blue{EL(p,L^c,i)}$ but it now consists, \blue{ for each commodity $c \in \mathcal{C}$, path $p \in P_{ij}^c$, and lifetime $\ell \in \{1,\ldots,EL(p,L^c,i\}$, of a single component $a^{(c,p,\ell)}_{sch_{ij}}(t) = \bigl[F^{(c,p,\ell)}\bigr]$ which is interpreted as the amount of packets to be forwarded, chosen from lowest to highest \textit{effective lifetime} $\ell$}. 
\end{itemize}

The resulting time complexity for the scheduler is:

\begin{itemize}
    \item The time to apply the action,  based on Algorithm \blue{\ref{alg:marl_el_scheduler_lelf}}, is $\bigO(|q_{i}|*|\mathcal{P}|*(\max_{p \in \mathcal{P}} \blue{EL(p,L_{max},i)}))$.
\end{itemize}
}
Note that since in this strategy, only the Router is MLP-based while Schedulers are rule-based, there is a significant improvement in training complexity. In addition, there is also an improvement in inference time complexity with respect to the strategy proposed in \ref{sec:send_max}.

\begin{algorithm}
	\caption{MARL EL LELF -- Scheduler}
	\begin{algorithmic}[1]
        \State $res\_flow \leftarrow C_{ij}$
        \State $max\_el = max\{\blue{EL(p,L^c,i)},\forall c\in \mathcal{C}, \forall p\in P_{ij}^c\}$
        \While {$res\_flow > 0 $}
            \For{$ \ell \in \{ 1 \ldots max\_el \} $}
                \For { \blue{$c \in \mathcal{C}$}}
                    \For {$p \in P_{ij}^{c}$}
                        \State $F = q_{i}.deque(c,p,\ell,1)$
                        \State $f_{ij}^{(c, p,\ell)} \leftarrow f_{ij}^{(c, p,\ell)} +F$
                        \State $res\_flow = res\_flow - F$
                    \EndFor
                \EndFor
			\EndFor
        \EndWhile
	\end{algorithmic}
    \label{alg:marl_el_scheduler_lelf}
\end{algorithm}

\subsection{Rule-based Strategy: Minimum Weight Router and LELF }
\label{sec:mec_lelf}
In order to understand the benefit of using an MLP-based router, we also test a strategy that combines LELF \textbf{Schedulers} as proposed in Section \ref{sec:lelf_heuristic} with a rule-based minimum weight \textbf{Router}. In order to assign packets to paths, the router updates the weights of the links based on the congestion of the queues of the respective paths, weighting the packets proportionally to their effective lifetime. Then, it proceeds with assigning the packets for each commodity to the minimum-weight path, up to its maximum capacity, assigning the remaining ones to the increasingly higher weight paths following the same criterion until exhaustion. 
The resulting time complexity of the Router referred to the Algorithm  \ref{alg:heuristic_mcr} is $\bigO(2*(|\mathcal{P}|*|\mathcal{E}|)+|\mathcal{P}|)$.

\begin{algorithm}
	\caption{\blue{MWR EL LELF} -- Router}
	\begin{algorithmic}[1]
        \State \texttt{\textbf{Assign weight to edges:}} Assign each edge $ij \in \mathcal{E}$ a weight $W_{ij}(t)$ equal to $\sum_{c\in\mathcal{C}}\sum_{p\in P_{ij}^c} \sum_{\ell \in \blue{1\ldots L^c}}|q_{i}^{(c, p,\ell)}(t)|*(abs(\ell -\blue{EL(p,\ell,i)})+1)$
        \State \texttt{\textbf{Calculate the weight for each Path:}} Assigns each path $p\in \mathcal{P}$ a weight $w_{p}$.
        \State \texttt{\textbf{Assign packets to paths:}} For each commodity $c\in \mathcal{C}$ assigns the newly arrived packets $b^{c}(t)$ to the best path up to its maximum capacity, then moves to the next one.
	\end{algorithmic}
    \label{alg:heuristic_mcr}
\end{algorithm}

\subsection{\blue{Neural Network Architecture Considerations}}
\label{sec:strategy_nn_improvements}
Despite the simplicity, the MLP-based neural network architectures are considered universal function approximators, as they can model continuous functions to an arbitrary degree of accuracy with sufficient parameters, making them a flexible choice for modeling complex relationships between the state and action spaces in RL environments. The choice of using an MLP architecture over others, like CNN, RNN, or even Transformers, is motivated by several factors like:
\begin{itemize}
    \item \textbf{Direct Mapping:} The MLP establishes a mapping of input on output with a low number of parameters, without initial/intermediate operations like in a Convolutional or Recurrent network, making it the most fitted to investigate the relationship between different state/action space designs.
    \item \textbf{Transparent operations:} The MLP architecture is simple and has no task-specific operations, like zero padding (CNN on images-like data), positional encoding (RNN sequential data), or attention mechanism (Transformer on sequential data). Considering also the simulation-based training/improvement environments and the distributed operational context, the MLP-based architecture is a convenient choice for computational efficiency in both the training and inference phases.
    \item \textbf{Efficient Implementation:} At its core, the MLP is based on matrix multiplication between layers, which are highly optimized operations in modern hardware, making them suitable for distributed operations even on low-resourced devices.
\end{itemize}
These factors positively impact the use of an MLP-based architecture in practice. In fact, it favors low latency even in a context with limited resources, thanks to optimizations of modern hardware and its low memory footprint due to the limited number of parameters. Hence, these features make it quickly adaptable through centralized or distributed training/refinement processes. This concept is even more true when compared to architectures like Transformer, which makes error tracking nearly impossible due to its extreme complexity.
A valid alternative to consider for further development is the Recurrent Neural Network (RNN). This network, which can model sequential data by capturing patterns that evolve over time, could improve performance at the cost of increased complexity and reduced operational clarity.

\section{Experimental Setting and Results}
\label{sec:expermiental_setting_and_results}

\begin{figure}
    \centering    \includegraphics[width=0.8\linewidth]{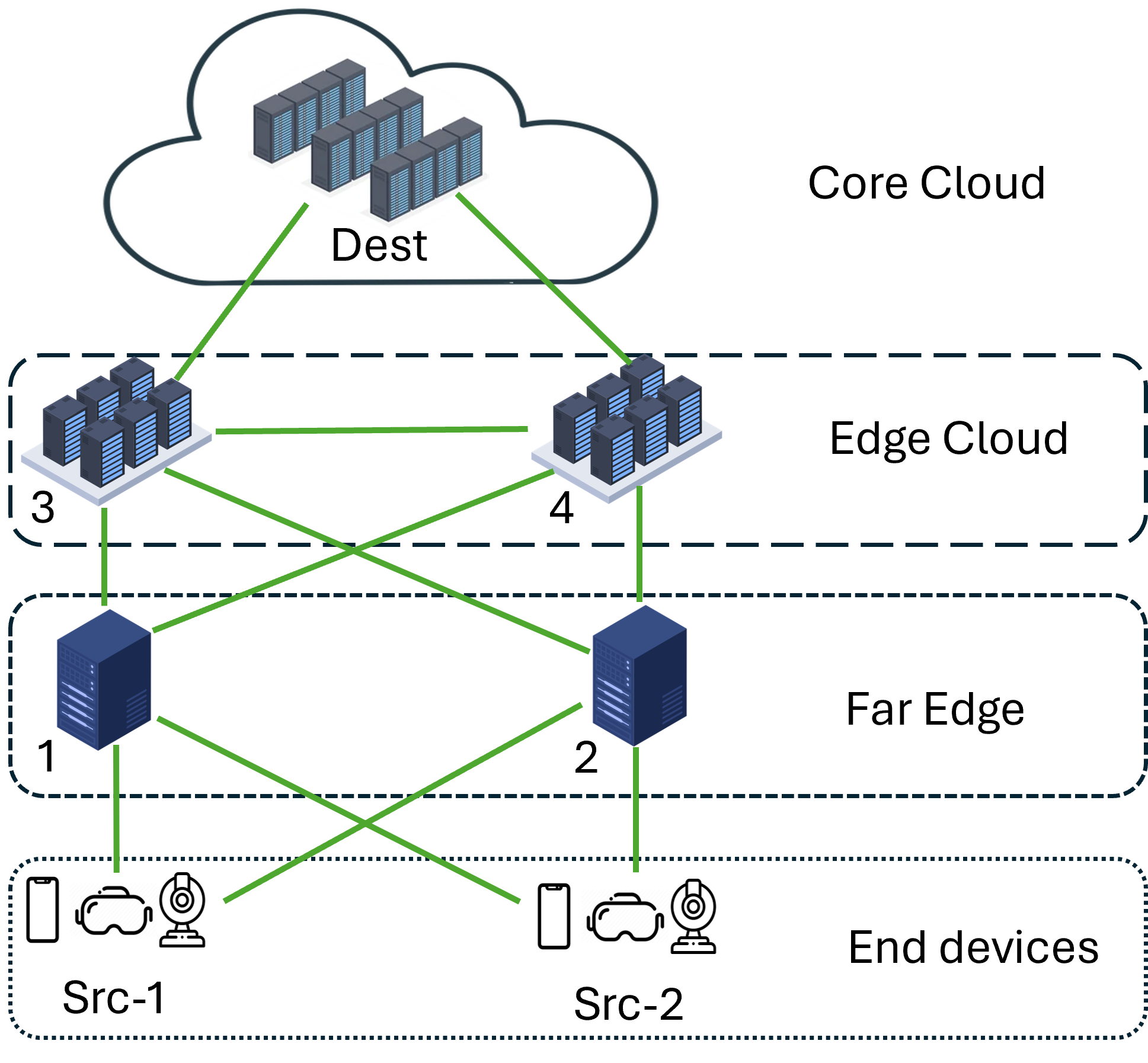}
    \caption{\small{The considered network scenario. End devices represent the source of information flow, while Core Cloud data centers represent the destination of information flows.}}
    \label{fig:unina_network}
    \vspace{-0.3cm}
\end{figure}

\begin{figure*}[!htb]
    \centering
    \includegraphics[width=0.99\linewidth]{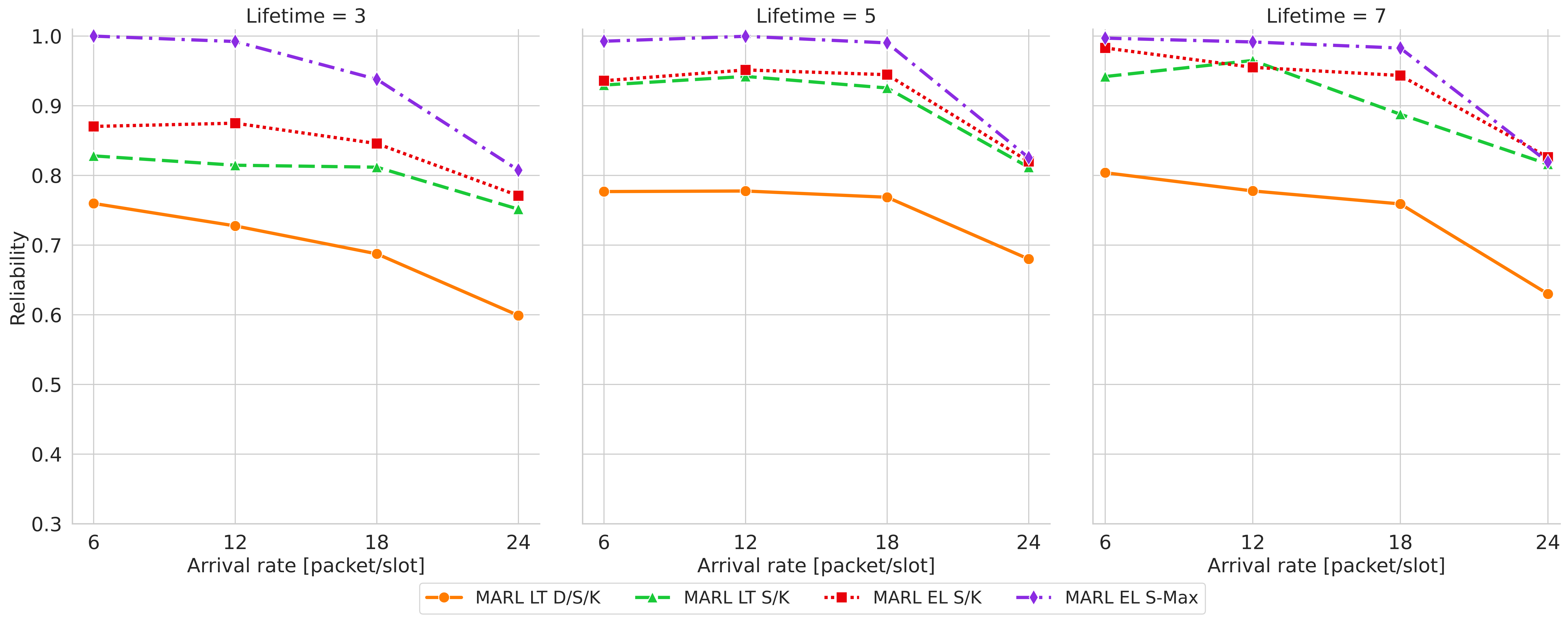}
    \caption{\small{Reliability averaged over 500 test episodes as a function of the arrival rate. Packets are generated at source nodes according to a Poisson distribution with a maximum lifetime 3 (on the left), 5 (center) and 7 (on the right). EL=Effective Lifetime, LT = Lifetime, D/S/K = Drop/Send/Keep}}
    \label{fig:unina_results}
\end{figure*}

We evaluate the proposed MADRL framework through simulations on a 7-node network topology reported in Figure \ref{fig:unina_network}, with a single centralized routing agent and $|\mathcal{E}|$ scheduling agents, one for each link/interface. The evaluation process includes two learning phases followed by a final testing phase. First, a standard training phase initializes the agents' policies. Next, a shorter improvement phase further refines these policies, with optimization parameters reset to enhance adaptability. Finally, a testing phase assesses the performance of the optimized policies.

\subsection{Considered Baseline}
\label{sec:umw_baseline}
\blue{To benchmark against state-of-the-art stochastic optimization methods, we implement a baseline using the Universal Max Weight (UMW)\cite{sinha2017optimal} algorithm, which we adapt for lifetime-constrained environments. UMW is a throughput-optimal dynamic control scheme that handles unicast, broadcast, multicast, and anycast traffic. It ensures stability and reduces latency by coupling min-cost routing with max-weight link scheduling, using virtual queues to relax precedence constraints.}
\blue{In this framework, the router pre-computes acyclic, virtual-queue-weighted min-cost paths at the source, while the distributed schedulers employ a simple First-In, First-Out (FIFO) policy. Our implementation accommodates the introduced lifetime constraints by considering packets expired if they are not delivered within their time limit.}

\subsection{Training Procedure Specifications}
The training process consists of an initial \textit{training phase} lasting 10,000 episodes, followed by a 4,000-episode \textit{improvement phase}, both using an initial learning rate of 0.001 \blue{and a batch size of 500 episodes. In both phases, agents employ an $\epsilon$-greedy strategy, with $\epsilon$ initialized at 1.0 and a decay factor of 0.95 applied after each episode, thereby ensuring a gradual shift from exploration to exploitation.} Each episode spans 50 time steps, after which the network resets and queues are cleared, allowing agents to explore a broad range of states. Each RL-based agent employs a Multi-Layer Perceptron (MLP) architecture for both its actor and critic networks, optimized with a Root Mean Square Error (RMSE) loss function. The Adam optimizer is consistently applied across both training phases. Each MLP has two hidden layers with respectively $m_{1}=128$ and $m_{2}=64$ hidden neurons.

\subsection{Reliability Results}
\label{sec:numerical_results}

\begin{figure*}
    \centering
    \includegraphics[width=0.99\linewidth]{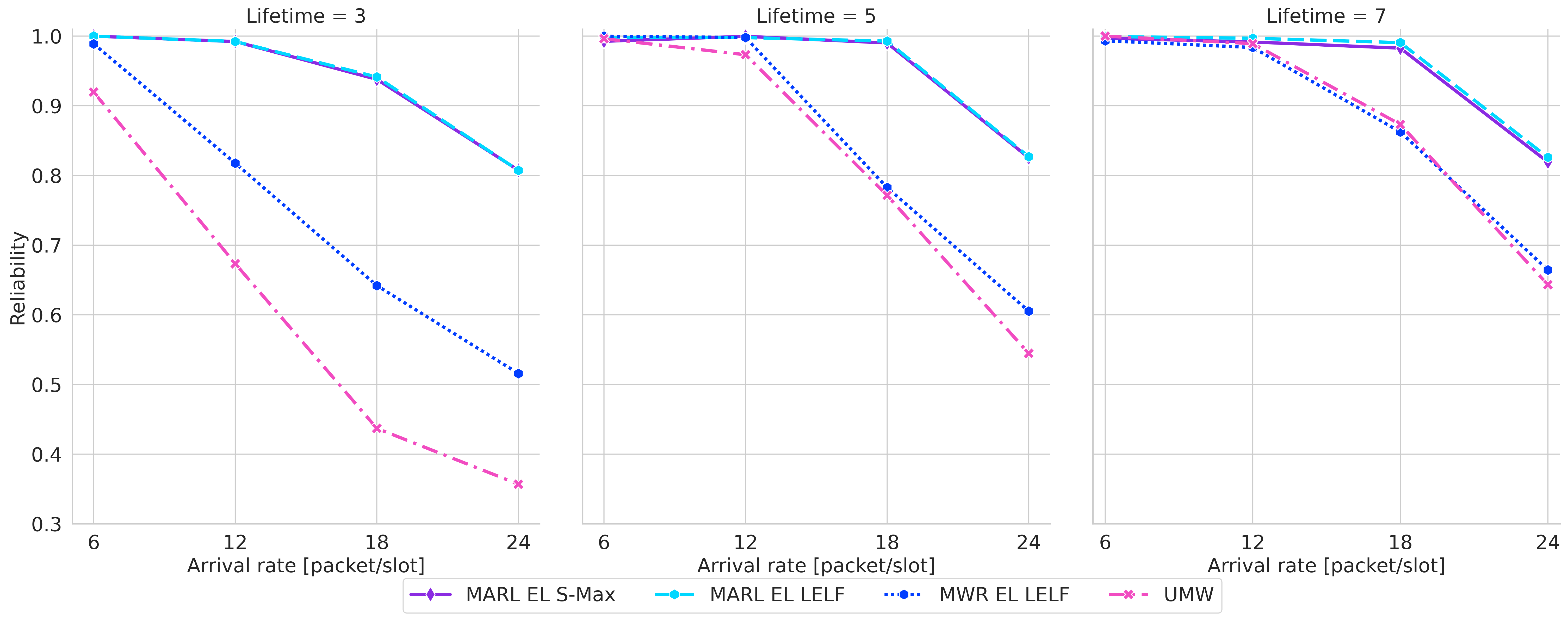}
    \caption{\small{Reliability averaged over 500 test episodes as a function of the arrival rate (The higher the better). Packets are generated at source nodes according to a Poisson distribution with a maximum lifetime 3 (on the left), 5 (center), and 7 (on the right). EL=Effective Lifetime, LT = Lifetime, D/S/K = Drop/Send/Keep}}
    \label{fig:thru_bestrl_rulebased}
\end{figure*}

\begin{figure*}
    \centering
    \includegraphics[width=\linewidth]{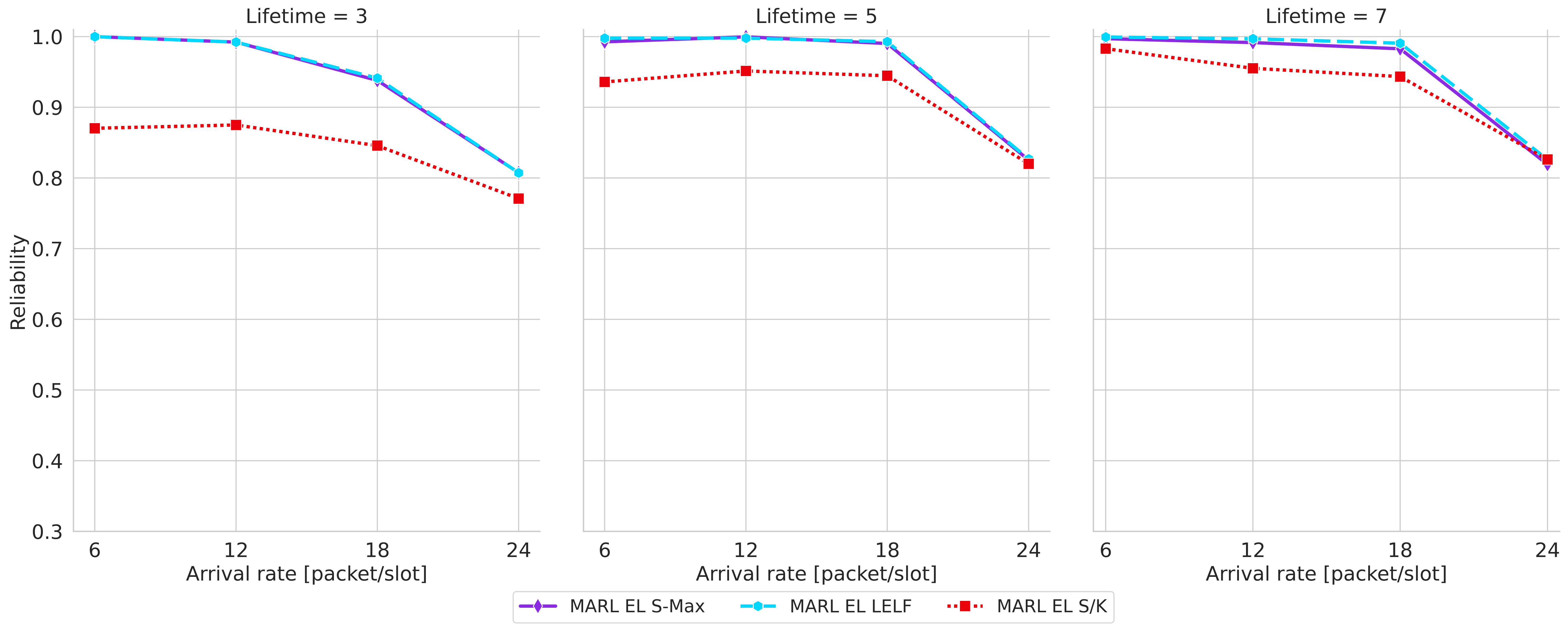}
    \caption{\small{Reliability averaged over 500 test episodes as a function of the arrival rate for the three best-performing strategies. Packets are generated at source nodes according to a Poisson distribution with a maximum lifetime 3 (on the left), 5 (center) and 7 (on the right). EL=Effective Lifetime, LT = Lifetime, D/S/K = Drop/Send/Keep}}
    \label{fig:thur_vs_compl}
\end{figure*}

\blue{In this section, we compare the performance of the previously introduced strategies, namely:}

\begin{itemize}
    \item \blue{MARL LT D/S/K introduced in section \ref{sec:base_maddpg}};
    \item  \blue{MARL LT S/K introduced in section \ref{sec:no_drop}};
    \item  \blue{MARL EL S/K introduced in section \ref{sec:effective_lifetime}};
    \item  \blue{MARL EL S-Max introduced in section \ref{sec:send_max}};
    \item \blue{MARL EL LELF introduced in section \ref{sec:lelf_heuristic}};
    \item  \blue{MWR EL LELF introduced in section \ref{sec:mec_lelf}};
    \item \blue{UMW baseline introduced in section \ref{sec:umw_baseline}}.
\end{itemize}

In the simulations based on the communication network reported in Figure \ref{fig:unina_network}, link capacities are fixed to 10 packets per time slot, and packets are generated at commodities' sources according to a \textit{Poisson} distribution with \textit{Arrival Rate} 3/6/9/12, resulting in aggregate arrival rates of 6/12/18/24. The initial maximum packets' lifetime is set to either 3,5, or 7. Throughout the training, improvement and test phases, both commodities have the same arrival rate $\bar b^c$ and lifetime $L^c$.
To make the results comparable across different lifetime and arrival rate configurations, reliability will be used in place of timely throughput, calculated as:
\begin{equation*}
Reliability=\frac{\texttt{timely throughput}}{\texttt{aggregate arr. rate}}  
\end{equation*}

The reliability performance of the proposed strategies were evaluated with a test phase consisting of 500 episodes and 50 steps per episode.
In Figure \ref{fig:unina_results}, we report the reliability performance measured for the first four proposed strategies in which each agent, either router or scheduler, is based on a neural network. These results highlight how the gradual design-driven dimensionality reduction in the action and state spaces leads to a gradual and significant increase in performance. In fact, moving from the MARL LT D/S/K strategy (orange line) to the MARL LT S/K strategy (green line) from which the drop action is removed, there is a significant increase in performance. Similarly, we observe a gradual increase in performance in the transition from MARL LT S/K (green line) to MARL EL S/K (red line) in which the effective lifetime (see equation \ref{eq:effective_lifetime}) is introduced. Finally, there was a significant increase in performance when we introduced the MARL EL S-Max strategy (purple line), which leverages the effective lifetime and removes the keep action from schedulers' action space, turning out as the best-performing strategy among the four considered.
Although the MARL EL S-MAX strategy performs best, it is more complex than the MARL EL S/K. We will discuss this tradeoff in section \ref{sec:complexity_analysis}.

Then, in Figure \ref{fig:thru_bestrl_rulebased}, we compare the MARL EL S-MAX with the baseline (UMW) and the strategies based on the LELF scheduling policy, namely the MARL EL LELF and the MWR EL LELF.
When comparing MARL EL S-MAX with the MARL EL LELF (light blue line), the MARL EL LELF exhibits better performance, mainly with larger Lifetime values. In addition, the MARL EL LELF has the advantage of a lower inference time complexity concerning schedulers' operations. 
Compared to the UMW (pink line) and the MWR+LELF (dark blue line), both MARL EL LELF and MARL EL S-MAX provide the best performance in all the considered settings, especially in the most challenging one with initial lifetime 3, while in the remaining scenarios (with initial lifetime 5 and 7), the performances are better (aggregate arrival rate $> 12$) or comparable (aggregate arrival rate $\leq 12$).
This analysis highlights the tradeoffs between reliability performance and complexity across the strategies, underscoring the practical advantages of MARL EL LELF in terms of its balance between performance and operational efficiency.

In Figure \ref{fig:thur_vs_compl}, we take the performance comparison further by analyzing the MARL EL LELF strategy in detail alongside the MARL EL S-MAX and MARL EL S/K strategies to highlight the differences in performance and complexity. The results demonstrate that the MARL EL LELF (light blue line) strategy not only achieves better reliability performance compared to MARL EL S-MAX (purple line) but also significantly reduces the inference time complexity due to its streamlined scheduler operations. However, it is worth noting that while the inference time complexity of MARL EL LELF is lower than that of MARL EL S-MAX, it is slightly higher or, at best, comparable, to that of the MARL EL S/K (red line) strategy, which benefits from a simpler algorithm; this aspect will be further analyzed in section \ref{sec:complexity_analysis}.

Notably, RL-based approaches are very effective, especially in the most challenging scenario, i.e., the lifetime 3 scenario, where the Effective Lifetime S-Max strategy achieves reliability almost comparable to the less challenging lifetime 7 scenario.

\begin{figure}[!hb]
    \centering
    \includegraphics[width=0.99\linewidth] {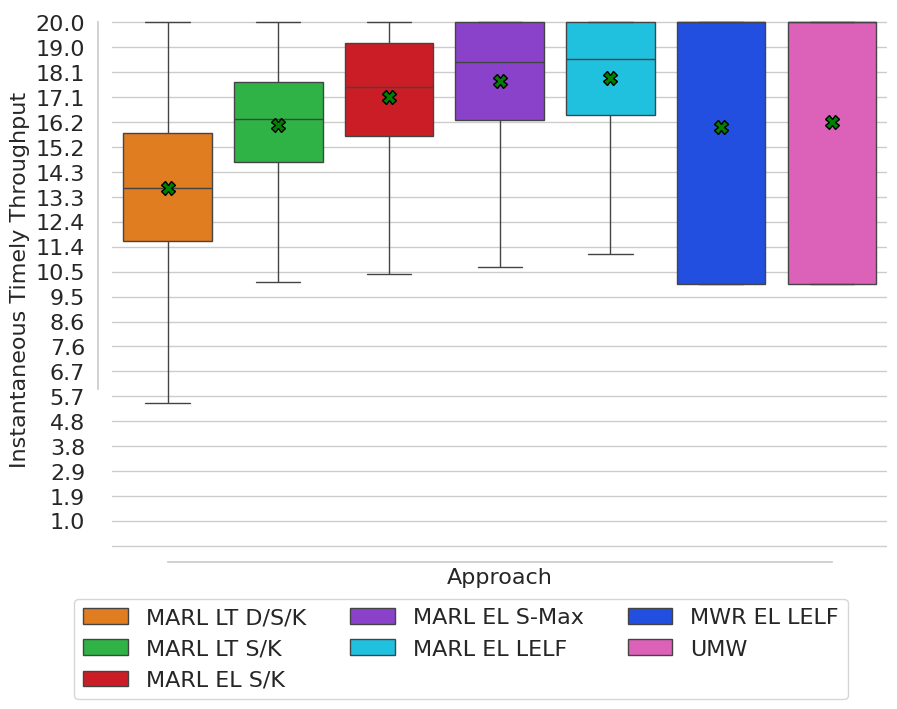}
    \caption{\small{Instantaneous Timely Throughput boxplot for lifetime 7 and aggregate arrival rate 18, showing the robustness of the proposed strategies. The green X indicates the average instantaneous timely throughput.}}
    \label{fig:enter-label}
\end{figure}

In addition to the improved average performance shown by the proposed strategies, we note that, especially in non-challenging conditions, namely with a high lifetime, the RL-based approaches prove to be more robust, as shown in Figure \ref{fig:enter-label} where a boxplot of the instantaneous timely throughput is reported. Even the MARL LT D/S/K approach shows a more predictable and stable behavior than the non-RL methods. 

\subsection{Inference Time Complexity Analysis Towards Sensitivity Analysis}
\label{sec:complexity_analysis}
The time complexity of the proposed strategies can be analyzed from two distinct perspectives: the complexity of the training process (when applicable) and the inference time complexity of agent interactions with the network.

The training complexity, in relation to the first aspect, is strongly impacted by the execution times of the simulation, which directly dictate the overall training duration. Moreover, under comparable conditions, convergence times are substantially influenced by the dimensionality of the action and state spaces. This underscores the pivotal importance of well-informed agent design in mitigating training overheads and enhancing overall performance.

To evaluate the complexity of agent interactions with the network, it is necessary to consider two components for each reinforcement learning (RL) agent. The first component is the computational time required to generate an action $a(t)$ based on the state $s(t)$. This depends on the architecture of the employed neural network (a multi-layer perceptron, MLP, in this case), as well as the sizes of the state and action spaces. The second component is the computational cost of executing the algorithm that translates the generated action into an effect on the environment. It is noteworthy that, for rule-based algorithms, the first component is not relevant.
The subsequent sections provide an analysis of the training complexity for each strategy, followed by a detailed examination of the per-agent complexity.

\begin{table*}[!htb]
    \centering
    \begin{tabular}{c|c|c}
        \textbf{Strategy} & \textbf{Observation} & \textbf{Action}\\
        \bottomrule
        \bottomrule
        MARL LT D/S/K & $\sum_{c\in\mathcal{C}} (L^c*|P_{ij}^{c}|)$ &  $\sum_{c\in\mathbf{C}} (3*L^c*|P_{ij}^{c}|)$\\
        MARL LT S/K & '' & $\sum_{c\in\mathbf{C}} (2*L^c*|P_{ij}^{c}|)$ \\
        MARL EL S/K & $\sum_{c\in\mathbf{C}}\sum_{p\in P_{ij}^c}\blue{EL(p,L^c,i)}$ & $\sum_{c\in\mathbf{C}}\sum_{p\in P_{ij}^c} (2*\blue{EL(p,L^c,i)})$\\
        MARL EL S-MAX & '' & $\sum_{c\in\mathbf{C}}\sum_{p\in P_{ij}^c} \blue{EL(p,L^c,i)}$ \\
        MARL EL LELF & '' & $\sum_{c\in\mathbf{C}}\sum_{p\in P_{ij}^c} \blue{EL(p,L^c,i)}$ \\
        \blue{MWR EL LELF} & '' & $\sum_{c\in\mathbf{C}}\sum_{p\in P_{ij}^c} \blue{EL(p,L^c,i)}$\\
        \bottomrule
        \bottomrule
    \end{tabular}
    \caption{In this table are reported the size of schedulers' state and action spaces grouped by proposed strategy. The order of the rows follows the incremental improvements proposed in the section, highlighting how the agents' design choices have led to a progressive and significant reduction in the size of the action and state spaces.}
    \label{tab:schedulers_action_observation_table_synthesis}
\end{table*}

\begin{table*}[!h]
    \centering
    \begin{tabular}{c|c|c}
        \textbf{Strategy} & \textbf{MLP} & \textbf{Algorithm}\\
        \bottomrule
        \bottomrule
        MARL LT D/S/K & $\bigO((\sum_{c\in\mathcal{C}} (L^c*|P_{ij}^{c}|))\ldots(\sum_{c\in\mathbf{C}} (3*L^c*|P_{ij}^{c}|)))$ & $\bigO(|\mathcal{P}|*|\mathcal{L}|*3)$ \\
        MARL LT S/K & $\bigO((\sum_{c\in\mathcal{C}} (L^c*|P_{ij}^{c}|))\ldots(\sum_{c\in\mathbf{C}} (2*L^c*|P_{ij}^{c}|)))$ & $\bigO(|\mathcal{P}|*|\mathcal{L}|*2)$ \\
        \textbf{MARL EL S/K} & $\bigO( (\sum_{c\in\mathbf{C}}\sum_{p\in P_{ij}^c}\blue{EL(p,L^c,i)})\ldots \sum_{c\in\mathbf{C}}\sum_{p\in P_{ij}^c} (2*\blue{EL(p,L^c,i)}))$  & $\bigO(|\mathcal{P}|*(\max_{p \in \mathcal{P}} \blue{EL(p,L_{max},i)})*2)$ \\
        \textbf{MARL EL S-MAX} & $\bigO( (\sum_{c\in\mathbf{C}}\sum_{p\in P_{ij}^c}\blue{EL(p,L^c,i)})\ldots(\sum_{c\in\mathbf{C}}\sum_{p\in P_{ij}^c} \blue{EL(p,L^c,i)}))$ & $\bigO(|q_{i}|*|\mathcal{P}|*(\max_{p \in \mathcal{P}} \blue{EL(p,L_{max},i)}))$  \\
        \textbf{MARL EL LELF} & -- & $\bigO(|q_{i}|*|\mathcal{P}|*(\max_{p \in \mathcal{P}} \blue{EL(p,L_{max},i)}))$ \\
        \blue{MWR EL LELF} & -- & $\bigO(|q_{i}|*|\mathcal{P}|*(\max_{p \in \mathcal{P}} \blue{EL(p,L_{max},i)}))$ \\
        \bottomrule
        \bottomrule
    \end{tabular}
    \caption{This table reports the schedulers' inference time complexity for comparison. The strategies in bold are those showing the best tradeoff between performance and inference time complexity. For the sake of brevity, the intermediate part of MLP complexity $*m_{1}+m_{1}*m_{2}+m_{2}*$ has been replaced with "$\ldots$". The "-" symbol stands for not applicable.}
    \label{tab:complexity_table_synthesis}
\end{table*}

\subsubsection{Training}

\begin{figure}[!hb]
    \centering
    \includegraphics[width=0.99\linewidth]{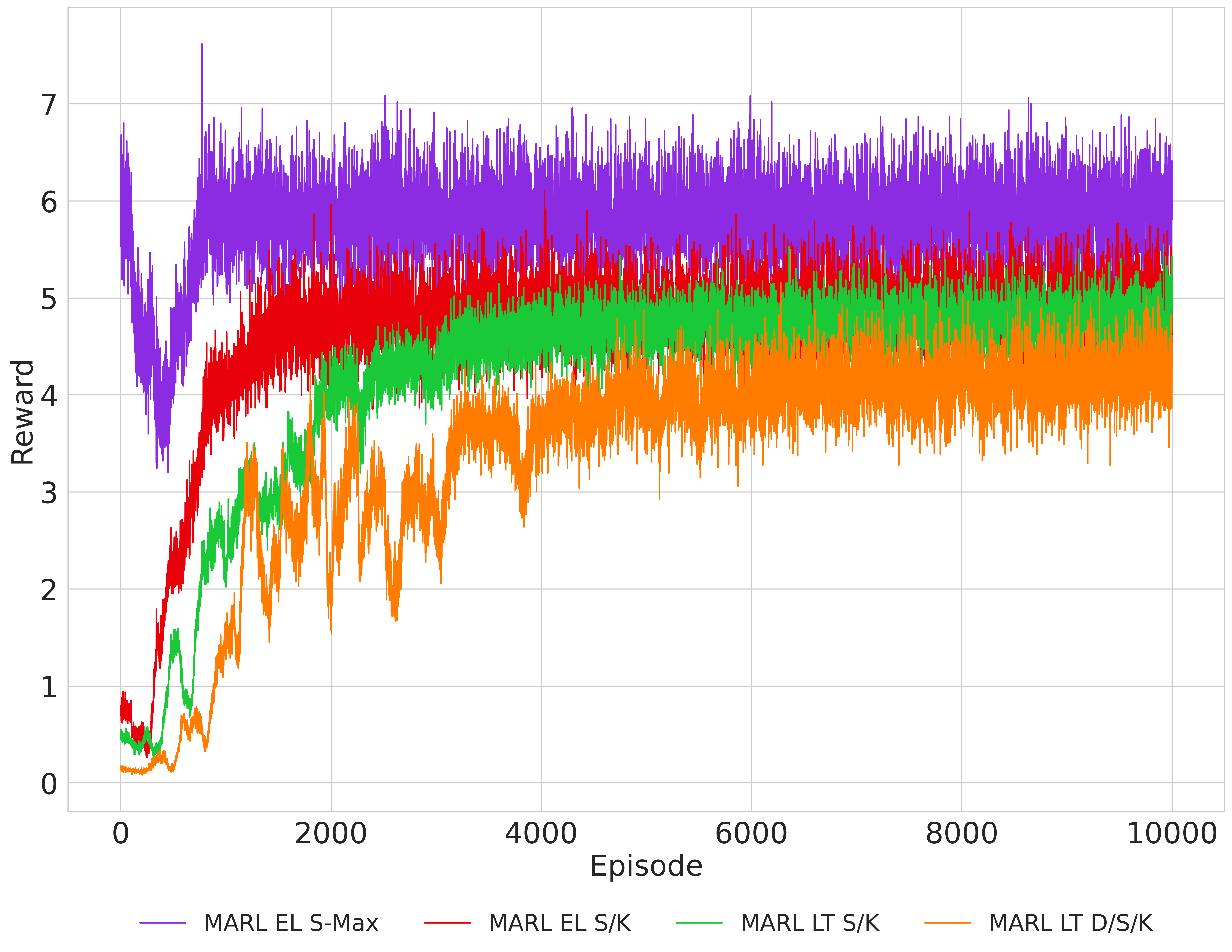}
    \caption{\small{Convergence plot of the reward over the 10 thousand training episodes, with initial lifetime 3 and aggregate arrival rate 6.}}
    \label{fig:convergence_plot_3}
\end{figure}

\begin{figure}[!hb]
    \centering
    \includegraphics[width=0.99\linewidth]{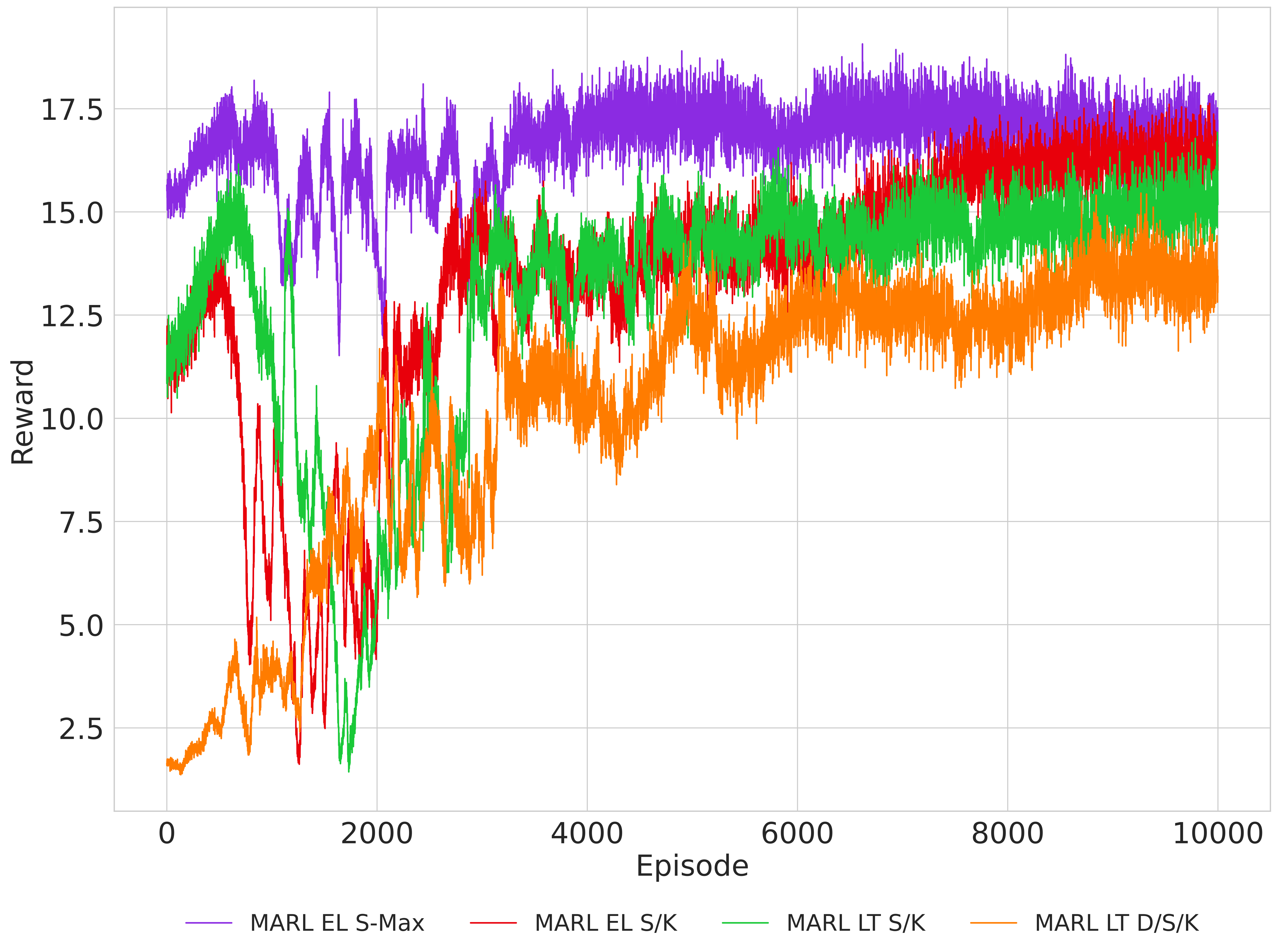}
    \caption{\small{Convergence plot of the reward over the 10 thousand training episodes, with initial lifetime 7 and aggregate arrival rate 18.}}
    \label{fig:convergence_plot_7}
\end{figure}

Figures \ref{fig:convergence_plot_3} and \ref{fig:convergence_plot_7} depict the reward convergence across 10,000 training episodes for the first four proposed strategies: MARL LT D/S/K, MARL LT S/K, MARL EL S/K, and MARL EL S-Max. The results underscore the effectiveness of the incremental improvements introduced in the proposed framework, particularly in addressing challenges related to training convergence. By systematically refining agent design based domain knowledge to optimize the size of the action and state spaces (as detailed in Table \ref{tab:schedulers_action_observation_table_synthesis}), these enhancements lead to faster and more stable convergence. As illustrated in the figures, the MARL EL S-Max strategy consistently outperforms the others, regardless of the initial lifetime or arrival rate. It is followed in performance by MARL EL S/K, MARL LT S/K, and, finally, MARL LT D/S/K, demonstrating the progressive benefits of the proposed design improvements. 
\blue{
It is worth emphasizing that the relatively high rewards observed at the beginning of training for MARL EL S-Max (Figures \ref{fig:convergence_plot_3} and \ref{fig:convergence_plot_7}) are intrinsic to the approach rather than artifacts of evaluation. This behavior stems from two main factors: the \emph{Effective Lifetime (EL)}, which binds each packet’s lifetime to its designated path and dynamically guides scheduling, and the scheduler strategy, which always saturates interface capacity and selects packets based on lifetime—a policy conjectured to be near-optimal. As a result, MARL EL S-Max initially outperforms strategies that must also learn packet dropping and rate allocation. When the network is empty, any path is effective and the router quickly attains high rewards; as congestion emerges, performance temporarily declines until it learns to consistently select efficient paths (after about 2,000 episodes). Early rewards are further destabilized by $\epsilon$-greedy exploration and the 500-episode buffer-filling period, with instability amplified in larger action spaces (Figure \ref{fig:convergence_plot_7}). Overall, the superior and more stable performance of MARL EL S-Max arises from the combined effects of EL, router learning, and a scheduling mechanism that is conjectured to be near-optimal among the strategies considered.}


\subsubsection{Router}
The router implementations utilized across the various strategies can be categorized into two distinct types.
The first type is an agent implemented using an MLP-based neural network, which is employed in the proposed MARL strategies. For this implementation, the temporal complexity associated with generating an action is given by $\bigO((|\mathcal{C}|+|\mathcal{V}|)*m_{1}+m_{1}*m_{2}+m_{2}*|\mathcal{P}|)$ where $(|\mathcal{C}|+|\mathcal{V}|)$ and $|\mathcal{P}|$ represent the sizes of the state and action spaces, and $m_{1}$ and $m_{2}$ denote the dimensions of the hidden layers in the neural network. The complexity associated with the algorithmic part is $\bigO(|\mathcal{P}|)$.

The second type of router is implemented in the MWR EL LELF strategy introduced in Section \ref{sec:mec_lelf}. This router's complexity, based on Algorithm \ref{alg:heuristic_mcr}, is $\bigO(2*(|\mathcal{P}|*|\mathcal{E}|)+|\mathcal{P}|)$.

\subsubsection{Scheduler}

Table \ref{tab:complexity_table_synthesis} summarizes the temporal complexities associated with each scheduler implementation. The results highlight how incremental improvements in the design of scheduling agents influence their inference time complexity.

The first improvement involves transitioning from the \textit{MARL LT D/S/K} strategy to \textit{MARL LT S/K}. This modification removes the drop action from the schedulers' action space, leading to a reduction in the inference time complexity. Although this reduction is constant, it has a substantial impact on performance, as highlighted by the reliability performance comparison reported in Figure \ref{fig:unina_results}. Consequently, the inference time complexity of \textit{MARL LT S/K} is strictly lower than that of \textit{MARL LT D/S/K}, i.e., \textit{MARL LT S/K} $<$ \textit{MARL LT D/S/K}.

The second improvement, transitioning from \textit{MARL LT S/K} to \textit{MARL EL S/K}, involves replacing the \textit{Lifetime} (LT) representation with an \textit{Effective Lifetime} (EL) representation in the network. This change substantially reduces the sizes of both the action and state spaces, thereby decreasing both components of the inference time complexity. Given that $\blue{EL(p,L^c,i)} \ll L^c \leq L_{max}, \forall c \in \mathcal{C} \texttt{ and } \forall p \in P^c$, we can assert that the inference time complexity of \textit{MARL EL S/K} is significantly lower than that of \textit{MARL LT S/K}, i.e., \textit{MARL EL S/K} $\ll$ \textit{MARL LT S/K}.

The third improvement, moving from \textit{MARL EL S/K} to \textit{MARL EL S-MAX}, eliminates the keep action from the schedulers' action space. This modification introduces multiple effects on the inference time complexity. While the reduction in the action space size decreases the complexity of the MLP component, the algorithmic component becomes more complex due to its dependency on the queue length, which has an upper bound of $|q{i}(t)|$. Despite this increased algorithmic complexity, this strategy achieves better performance compared to \textit{MARL EL S/K}. Therefore, the inference time complexity of \textit{MARL EL S-MAX} is greater than or equal to that of \textit{MARL EL S/K}, i.e., \textit{MARL EL S-MAX} $\gg$ \textit{MARL EL S/K}.

The fourth improvement, transitioning from \textit{MARL EL S-MAX} to \textit{MARL EL LELF}, introduces the \textit{LELF} scheduling policy, which is the only non-machine-learning-based strategy. In this case, the inference time complexity consists solely of the algorithmic component. As a result, the inference time complexity of \textit{MARL EL LELF} is significantly lower than that of \textit{MARL EL S-MAX}, i.e., \textit{MARL EL LELF} $\ll$ \textit{MARL EL S-MAX}. However, when compared to \textit{MARL EL S/K}, which has the lowest MLP complexity, the algorithmic component of \textit{MARL EL LELF} is bounded by the queue size, leading to an inference time complexity that is significantly greater than the algorithmic component of \textit{MARL EL S/K}, i.e., \textit{MARL EL LELF} $\ge$ \textit{MARL EL S/K}.

\section{Discussion and Future Directions}
\label{sec:discussion_and_future_dev}
The proposed MA-DRL framework combines centralized routing with distributed scheduling to dynamically allocate paths and manage packet transmissions. By leveraging the Multi-Agent Deep Deterministic Policy Gradient (MADDPG) technique, the framework effectively balances the complexity of learning with the performance requirements of latency-critical services. Our results demonstrate the ability of this approach to outperform traditional stochastic optimization-based methods, showcasing its potential for real-world network environments.

A critical aspect of the framework is its integration of networking domain knowledge into the design of reinforcement learning (RL) strategies. By incorporating rule-based policies where appropriate, the framework reduces the complexity of RL agents, enabling more efficient learning while maintaining high performance. This flexibility allows the framework to strike an optimal balance between data-driven decision-making and deterministic control, highlighting the importance of domain-specific insights in designing scalable and effective solutions.

Additionally, an inference time complexity analysis (towards a sensitivity analysis) and incremental strategy improvements conducted in this study provide valuable insights into the components of the state and action spaces most relevant to enhancing performance. These findings pave the way for further optimizations in agent design, enabling the development of tailored solutions for a variety of network control scenarios.

While this work provides a solid foundation for latency-critical network control using MA-DRL, several directions for future research can be pursued to extend its applicability and effectiveness:
\begin{itemize}
    \item Scalability and Robustness: Testing the framework on larger networks and under dynamic conditions, such as fluctuating link states or traffic surges.
    \item Energy Efficiency: Integrating energy-aware policies to balance latency and resource consumption.
    \item Alternative Architectures: Exploring advanced DRL techniques and other neural network architectures such as Recurrent Neural Networks and Graph Neural Networks.
    \item Real-World Validation: Deploying the framework in testbeds or emulators to evaluate its practicality and real-world impact.
\end{itemize}

\section{Conclusion}
\label{sec:conclusions}
This work addresses the critical challenge of achieving efficient and reliable network control for latency-sensitive applications by introducing a novel Multi-Agent Deep Reinforcement Learning (MADRL) framework. By modeling the Delay-Constrained Maximum-Throughput (DCMT) network control problem as a Markov Decision Process (MDP), we establish a foundation for deriving optimal routing and scheduling policies to maximize timely packet delivery.

The proposed MADRL framework integrates centralized routing and distributed scheduling agents, enabling a collaborative learning process that balances complexity and performance. By progressively incorporating networking domain knowledge, our approach reduces the decision burden on reinforcement learning (RL) agents, allowing rule-based policies to handle tasks more suited to deterministic control. This hybrid design significantly enhances the framework's efficiency and practicality in addressing real-world network control challenges.

Furthermore, the incremental improvements in strategy design offer valuable insights into the impact of action and state space dimensions on agent performance and convergence behavior. These findings underscore the importance of informed agent design in optimizing performance for latency-critical services.

This work aims to contribute not only to advancing the state-of-the-art in DRL-based network control but also to provide a flexible and extensible framework for addressing broader challenges in dynamic and heterogeneous network environments. Future research can build on these findings to further enhance scalability, robustness, and applicability across diverse use cases, solidifying the role of MADRL as a cornerstone for next-generation network control solutions.

\section*{Acknowledgements}
This work was partially supported by the European Union under the Italian National Recovery and Resilience Plan (NRRP) of NextGenerationEU, partnership on "Telecommunications of the Future" (PE00000001 - program "RESTART"), by the PRIN project "Resilient delivery of real-time interactive services over NextG compute-dense mobile networks" (E53D2300055000), and by funds from the US National Science Foundation as specified in the RINGS program.





%

\bibliographystyle{IEEEtran}
\bibliography{biblio}
\end{document}